\documentclass[%
 reprint,
superscriptaddress,
 amsmath,amssymb,
 aps,
]{revtex4-2}

\usepackage{graphicx}
\usepackage{dcolumn}
\usepackage{bm}

\begin{document}

\preprint{APS/123-QED}

\title{Flow and clogging of capillary droplets}

\author{Yuxuan Cheng}\email{yuxuan.cheng@yale.edu}
 \affiliation{Department of Physics, Yale University, New Haven, Connecticut, 06520, USA.}

\author{Benjamin F. Lonial}
\affiliation{
Department of Physics, Emory University, Atlanta, GA 30322, USA.
}
\author{Shivnag Sista}
 \affiliation{Department of Physics, Yale University, New Haven, Connecticut, 06520, USA.}

\author{David J. Meer}
\affiliation{
Department of Physics, Emory University, Atlanta, GA 30322, USA.
}

\author{Anisa Hofert}
\affiliation{
Department of Physics, Emory University, Atlanta, GA 30322, USA.
}

\author{Eric R. Weeks}
\affiliation{
Department of Physics, Emory University, Atlanta, GA 30322, USA.
}
\author{Mark D. Shattuck}
\affiliation{
Benjamin Levich Institute and Physics Department, The City College of New York, New York, New York 10031, USA.
}
\author{Corey S. O'Hern}\email{corey.ohern@yale.edu}
 \affiliation{Department of Physics, Yale University, New Haven, Connecticut, 06520, USA.}
\affiliation{
Department of Mechanical Engineering and Materials Science, Yale University, New Haven, Connecticut, 06520, USA.
}

\affiliation{
Program in Computational Biology and Bioinformatics, Yale University, New Haven, Connecticut, 06520, USA.
}

\date{\today}

\begin{abstract}
{Capillary droplets form due to surface tension when two immiscible fluids are mixed.  We describe the motion of gravity-driven capillary droplets flowing through narrow constrictions and obstacle arrays in both simulations and experiments.  Our new capillary deformable particle model recapitulates the shape and velocity of single oil droplets in water as they pass through narrow constrictions in microfluidic chambers. Using this experimentally validated model, we simulate the flow and clogging of single capillary droplets in narrow channels and obstacle arrays and find several important results. First, the capillary droplet speed profile is nonmonotonic as the droplet exits the narrow orifice, and we can tune the droplet properties so that the speed overshoots the terminal speed far from the constriction. Second, in obstacle arrays, we find that extremely deformable droplets can wrap around obstacles, which leads to decreased average droplet speed in the continuous flow regime and increased probability for clogging in the regime where permanent clogs form.  Third, the wrapping mechanism causes the clogging probability in obstacle arrays to become nonmonotonic with surface tension $\Gamma$.  At large $\Gamma$, the droplets are nearly rigid and the clogging probability is large since the droplets can not squeeze through the gaps between obstacles. With decreasing $\Gamma$, the clogging probability decreases as the droplets become more deformable.  However, in the small-$\Gamma$ limit, the clogging probability increases since the droplets are extremely deformable and wrap around the obstacles. The results from these studies are important for developing a predictive understanding of capillary droplet flows through complex and confined geometries.} \\
 
\end{abstract}

\maketitle


\section{Introduction}
\label{intro}
Capillary droplets, which consist of immiscible liquid droplets dispersed in another continuous phase, are ubiquitous in food science~\cite{CHUNG2014106,Bai2021}, the pharmaceutical industry~\cite{FREITAS200676,ALBERT2019302}, petroleum products~\cite{doi:10.1021/ie00050a014,UMAR2018673}, and pollution remediation~\cite{su132212339}. Being able to control how capillary droplets flow and interact with their surroundings is crucial for enhancing food emulsion stability, optimizing drug production and delivery, and improving oil recovery rates, while minimizing environmental risks associated with oil exploration. For example, the flow rate of capillary droplets can be tuned by varying the channel width and by designing obstacle arrays in microfluidic chambers~\cite{Bick2021,1,2,3,4,5}, cell sorting instruments~\cite{6,7,16}, and cell encapsulation devices~\cite{8,9,10}. However, we do not yet have a fundamental understanding of the dynamics of capillary droplets flowing through complex, confined geometries as a function of the droplet properties.

Flows of hard, frictional grains through narrow constrictions, such as hoppers, silos, and obstacle arrays, display complex spatiotemporal dynamics~\cite{PhysRevE.94.022901,Pascot2020,10.1063/1.3669495}. Over $50$ years ago, Beverloo and co-workers~\cite{BEVERLOO1961260} carried out experimental studies of hopper flows for a wide range of dry granular materials and proposed an empirical form for the flow rate $Q$ (in the continuous flow rate regime) versus the orifice width $w$: $Q \sim (w/\sigma_{\rm avg} -k)^{\beta}$, where $\sigma_{\rm avg}$ is the mean grain size, the flow arrests for $w/\sigma_{\rm avg} \lesssim k$, $k \gtrsim 1.5$ for rigid grains, and $\beta$ is the power-law scaling exponent that can be tuned by varying the ratio of the dissipation from particle-particle contacts and from the background fluid~\cite{Yuxuan}. In addition, researchers have found that placing an obstacle $\sim 3\sigma_{\rm avg}$ above the orifice can reduce clogging in silo flows of hard, frictional grains~\cite{Zuriguel,Lozano,Gao,Alessio2021,Gai,Gao2019,Alison}. Experimental and computational studies have also shown that the permeability of an obstacle array to granular flow can be tuned by varying the density, shape, and size of the obstacles~\cite{Reichhardt,Nguyen,ReichhardtC,Peter2018}. 

In contrast, for deformable particles, it is possible to obtain $Q >0$ with no clogging even for $w/\sigma_{\rm avg} < 1$. Unlike hard grains, deformable particles can change their shapes to squeeze through narrow constrictions that are much smaller than their undeformed diameters. However, there have been few studies of flows of deformable particles through narrow constrictions. Many experimental studies have instead focused on hydrogel particle flows in hoppers and silos with $w/\sigma_{\rm avg} \sim 2$-$3$, which do not give rise to large particle deformation. For example, recent studies have found that hydrogel particles with smaller elastic moduli are less likely to clog~\cite{Hong2017,Tao2021,D2SM00318J} and possess flow rates that are much faster than those for hard particles, such as glass beads~\cite{D0SM01887B}. These prior results for hydrogel particles do not address how particle flow rates are affected by large particle deformation. Thus, an important open question is: does significant particle shape change (i.e. by more than $50\%$) still enhance flow through narrow constrictions and other confined geometries?

The interplay between particle deformability and the geometry of obstacle arrays can give rise to a new physical mechanism, {\it particle wrapping}, which can strongly influence the dynamics of capillary droplets as they flow through obstacle arrays. When colliding with an obstacle, the highly deformable droplet can wrap around the obstacle with both ends of the droplet elongating in the flow direction until the energy penalty from the droplet surface tension matches the work done by the fluid driving force. At this point, both ends of the droplet become stationary, and then one end of the droplet moves in the opposite direction of the driving force, causing the particle to unwrap. Particle wrapping can even lead to single-particle clogging in obstacle arrays. In this case, the droplet is in force balance after wrapping around an obstacle, where the fluid forces on the droplet are supported by contact forces on the droplet from neighboring obstacles.  Thus, particle wrapping can lead to much longer transit times through obstacle arrays compared to those for hard particles with the same undeformed diameter. Despite its importance, the particle wrapping mechanism has not been studied in detail in droplet flows through obstacle arrays.

A key factor that contributes to the knowledge gap concerning flows of highly deformable particles in confined geometries is the lack of an accurate and efficient computational model for these flows. At submicron length scales, e.g. within microfluidic devices, inertial forces are typically much smaller than capillary forces generated by particle surface tension, which strongly influences the behavior of multi-phase fluid flows. Computational models for multi-phase fluid flows include grid-based Eulerian techniques, such as the volume-of-fluid~\cite{HIRT1981201} and Lattice Boltzmann methods~\cite{D3SM01648J,Coelho2023}, and particle-based Lagrangian techniques, such as the smoothed particle hydrodynamics method~\cite{BREINLINGER201314,Gingold,MonaghanJ}. Grid-based methods determine the amount of fluid and gas in each cell and calculate the surface energy by reconstructing the fluid-gas interface. However, this method does not strictly conserve mass, and the surface energy can depend on the interface reconstruction algorithm~\cite{HIRT1981201}. For particle-based methods, the multi-phase system is discretized into ``particles" with each particle assigned as gas or liquid with a specified mass. Therefore, interface tracking is not needed, and the multi-phase fluid behavior is governed by particle-particle interactions. However, it is challenging to determine {\it a priori} the particle-particle interactions that yield the correct contact angles at the multi-phase interfaces~\cite{Morris,BREINLINGER201314}. Particle-based methods are also computationally demanding since they simulate the droplet interiors in addition to the interfaces. As a result, most particle-based simulations of multi-phase fluids are limited to short time scales and small systems. 

In this article, we describe combined experimental and computational studies that enable the development of new efficient computational methodologies and provide new insights for flows of deformable particles through narrow constrictions and obstacle arrays. In the experimental studies, we measure the speed and shape of capillary droplets as they flow through narrow openings as a function of the orifice width and gravitational driving force. For the computational studies, we develop a new deformable particle (DP) model\cite{Boromand2018,Boromand2019,Wang2021} with surface tension (in two dimensions, 2D) that can be implemented to efficiently model thousands of capillary droplets flowing through complex, confined geometries. Varying surface tension in the experiments by more than a factor of $2$ is challenging, and thus we carry out simulations of extremely deformable particles with surface tensions that vary by more than a factor of $10^3$, which accentuates the particle wrapping mechanism. We first validate the DP simulations by comparing the droplet shape dynamics in experiments and simulations for a single droplet flowing through narrow constrictions. We then investigate the important physical parameters that control the flow rate and clogging propensity of single deformable particles moving through obstacle arrays using the experimentally validated DP simulations. {Through this article, we seek to encourage future experimental studies that  will investigate droplet squeezing and wrapping behavior of capillary droplets flowing through obstacle arrays.}
 
We find several important results. First, when a capillary droplet is pushed by gravity (or buoyancy) through a narrow constriction, its speed initially decreases as it enters the orifice and deforms. After it reaches a minimum speed, the droplet begins to speed up as it exits the orifice. However, as the speed increases, it can overshoot the droplet's terminal speed. We find that the amplitude of the overshoot decreases with increasing gravitational driving force (relative to the capillary forces). Second, we show that highly deformable droplets flow more slowly in obstacle arrays compared to droplets with larger surface tension, due to the wrapping mechanism. Thus, for capillary droplets flowing through obstacle arrays, the clogging probability, as characterized by the average distance $\lambda$ traveled by a droplet before it clogs, is non-monotonic as a function of surface tension. For small values of surface tension, $\lambda$ first increases with increasing surface tension, reaches a peak that depends on the obstacle spacing, and then decreases with continued increases in surface tension. In contrast, models of soft particles that do not model explicit changes in particle shape cannot capture particle wrapping\cite{Durian1995,Durian1997}, and thus cannot accurately model flows of capillary droplets moving through obstacle arrays in the low surface tension limit.

The remainder of the article is organized as follows. In Section~\ref{sim_methods}, we describe the computational methods including descriptions of the deformable particle model\cite{Boromand2018,Boromand2019,Wang2021} for droplets and the soft (semi-rigid) particle model\cite{Durian1995,Durian1997} and the equations of motion and boundary conditions for discrete element method (DEM) simulations of droplet flows through narrow channels and obstacle arrays. In Section~\ref{expmethods}, we introduce the experimental setup and method to generate capillary droplets and flows. In Section~\ref{results}, we describe the main results from the combined experimental and simulation studies. We first characterize the droplet speed and shape dynamics as a function of the distance $h$ from the orifice for both oil-in-water experiments and DPM simulations of droplet flows through narrow channels. We calibrate the results for the droplet speed and shape from the DP simulations to those from the experiments. Using the experimentally validated DP simulations, we quantify the nonmonotonic behavior in the droplet speed profile and the clogging probability of single droplets flowing through obstacle arrays as a function of the droplet surface tension. In Section~\ref{discussion}, we provide conclusions and propose future research directions, including studies of multi-droplet flows through obstacle arrays with droplet breakup and coalescence that can predict the steady-state droplet size distribution. We also include five Appendices to supplement the discussion in the main text. In Appendix A, we present experimental data for the near-wall droplet speed profile. In Appendix B, we investigate how the results of the DP simulations depend on the number of vertices $N_v$ in each deformable particle and the perimeter elasticity $K_l$. In Appendix C, we provide additional details concerning the definition of a clogging event and the clogging probability for droplets flowing through obstacle arrays. In Appendix D, we identify the surface tension regime for which the simulations can achieve continuous flows in obstacle arrays. In Appendix E, we quantify the dependence of the steady-state speed of the droplets on the near-wall drag coefficient.  

\section{Simulation Methods}
\label{sim_methods}

In this section, we describe the methods for simulating 2D gravity-driven flows of a single droplet through narrow channels and obstacle arrays.  We first illustrate the geometry of the narrow channels and obstacle arrays. We then introduce two models for describing the droplet shape and interactions with the walls: {1) the deformable particle (DP) model that treats each droplet as a polygon with $N_v$ vertices, employs a shape-energy function to penalize changes in droplet area and changes in the separations between vertices, and enforces excluded volume interactions using pairwise purely repulsive forces between the droplet vertices and walls} and 2) the soft particle (SP) model, which tracks the centers of mass of the droplets and enforces excluded volume interactions using pairwise purely repulsive forces that increase with the overlap between a droplet and the walls. For both models, we define the forces on the droplets that arise from the shape-energy function, droplet-wall interactions, and dissipative forces from the background fluid, and provide the equations of motion for the droplet trajectories. Finally, we describe the method for initializing the droplet positions and velocities in narrow channels and obstacle arrays.      

\subsection{Geometry for flows through narrow channels and obstacle arrays}
\label{geometry}
\begin{figure}[!h]
\centering
\includegraphics[width=0.95\columnwidth]{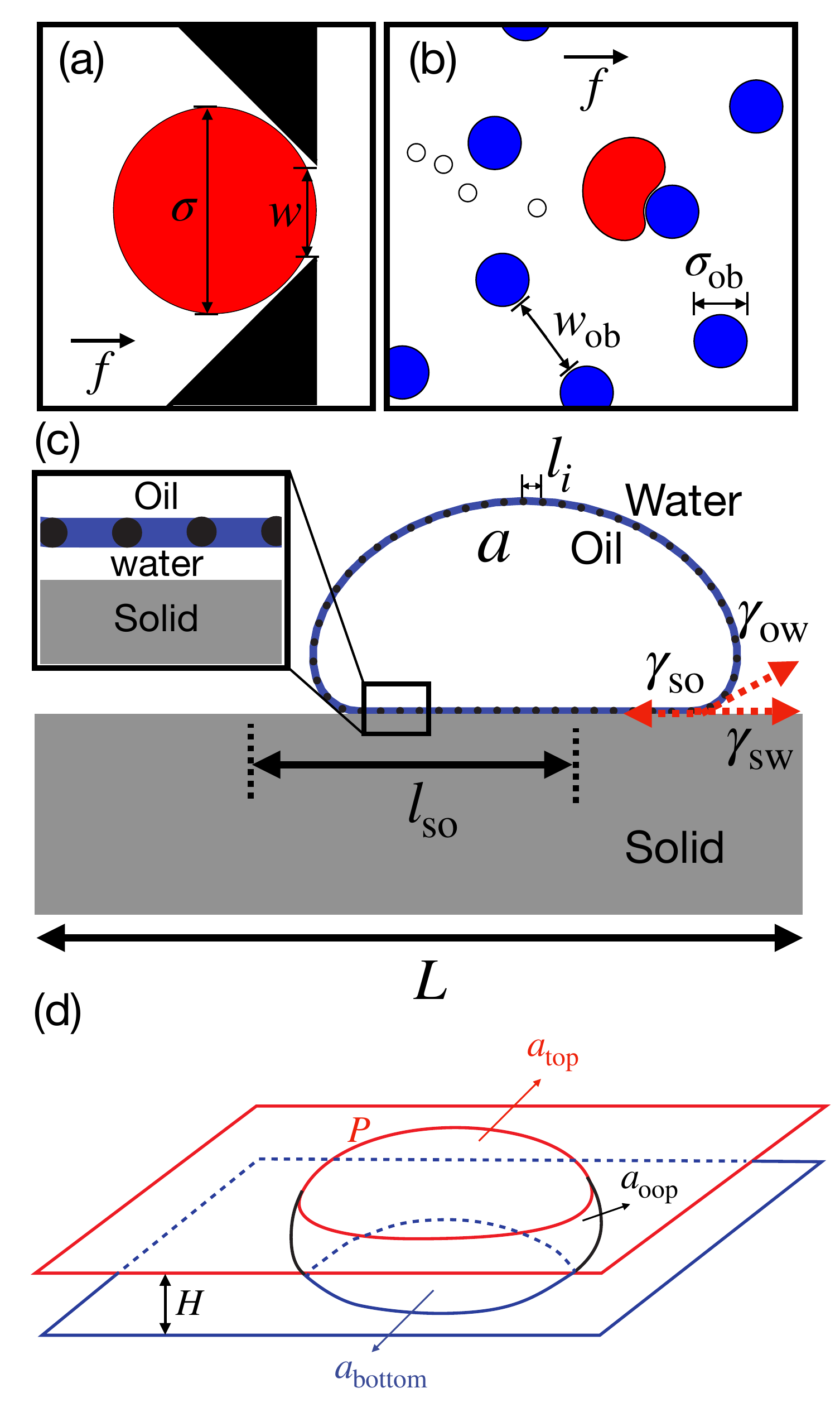}
\centering
\caption{Schematic of {the top view} of (a) a droplet (shaded red) with undeformed diameter $\sigma$ flowing through a narrow orifice (shaded black) with width $w=0.4 \sigma$ and (b) a droplet (shaded red) flowing through an obstacle array (shaded blue) with random obstacle positions, obstacle sizes $\sigma_{\rm ob} = 0.3\sigma$, and minimum separation $w_{\rm ob}=\sigma$. The open circles represent the droplet's trajectory.  The rightward pointing arrow indicates the direction of {droplet flow}. (c) Illustration of {the top view} of the DP model in 2D with solid-oil $\gamma_{\rm so}$, oil-water $\gamma_{\rm ow}$, and solid-water $\gamma_{\rm sw}$ line tensions. In the oil-in-water experiments, a thin layer of water exists between the oil droplets and chamber walls {as shown in the inset}, and thus $\gamma_{\rm so} = \gamma_{\rm ow} = \gamma_{\rm sw}$. $l_{\rm so}$ is the contact length between the solid wall and oil droplet, $L$ is the length of the solid wall, $l_i$ is the separation between adjacent droplet vertices, and $a$ is the droplet area. {(d) Schematic of the side view of a quasi-2D capillary droplet placed between two parallel walls with droplet thickness $H$, top-view 2D perimeter $P$, top-facing surface area $a_{\rm top}$, bottom-facing surface area $a_{\rm bottom}$,  and out-of-plane surface area $a_{\rm oop}$. For the DP model in 2D, $a = a_{\rm top} = a_{\rm bottom}$.}}
\label{fig:model}
\end{figure}

The narrow channels in the 2D simulations are constructed using two infinitely long wedges, one on the top and one on the bottom that form an orifice between them with width $w<\sigma$, where $\sigma$ is the diameter of the undeformed droplet. [See Fig.\ref{fig:model}(a).]  The obstacle arrays are located within $L \times L$ simulation boxes with periodic boundary conditions and $L = 500\sigma$. The circular obstacles with diameter $\sigma_{\rm ob}$ are placed inside the simulation box using a Poisson sampling procedure with separations $w>w_{\rm ob}$. [See Fig.\ref{fig:model}(b).] 

\subsection{Deformable particle model}

When droplets squeeze through narrow constrictions, they can experience dramatic shape changes. To efficiently describe large changes in shape that occur during gravity-driven droplet flows, we employ the recently developed deformable particle (DP) model~\cite{Boromand2018,Boromand2019,Wang2021}. In 2D, the droplets are modeled as deformable polygons with $N_v$ vertices. The shape-energy function $U_s$ for each deformable particle includes three terms: 
\begin{equation}
\label{shape_energy}
U_{s} = \frac{k_a}{2} (a-a_{0})^2 + \frac{k_l N_v}{2}\sum_{i=1}^{N_v} (l_{i}-l_{0})^2 + U_{\gamma}.
\end{equation}
The first term imposes a harmonic energy penalty for changes in the droplet area $a$ from the preferred value $a_{0}$ and $k_a$ controls the fluctuations in droplet area. The second term imposes a harmonic energy penalty for deviations in the separations $l_{i}$ between adjacent vertices $i$ and $i+1$ from the equilibrium length $l_{0}$ and $k_l$ controls fluctuations in the separations between adjacent vertices. This term ensures that vertices are distributed evenly on the droplet surface, preventing them from aggregating in the direction of gravity. The factor of $N_v$ in the numerator of the second term of Eq.~\ref{shape_energy} makes $U_s$ independent of $N_v$.  We characterize the shape of the droplet using the dimensionless shape parameter: ${\cal A} = (\sum_{i=1}^{N_v} l_i)^2/(4\pi a)$, where ${\cal A}=1$ for a circle. The third term gives the energy arising from line tension:
\begin{equation}
U_{\gamma} = \gamma_{\rm so} l_{\rm so} +  \gamma_{\rm sw} (L - l_{\rm so})+ \gamma_{\rm ow} l_{\rm ow},
\label{eq:surface}
\end{equation}
where $\gamma_{\rm so}$, $\gamma_{\rm sw}$, and $\gamma_{\rm ow}$ are the solid-oil, solid-water, and oil-water line tension coefficients. (See Fig.~\ref{fig:model} (c) for a schematic of the solid wall-oil droplet-water interface in 2D.) $l_{\rm so}$ and $l_{\rm ow}$ are the contact lengths at the oil-solid and oil-water interfaces, respectively. $L-l_{\rm sw}$ is the length of the solid wall that is not in contact with the oil droplet. In the oil-in-water experiments described here, thin layers of water exist between the oil droplets and chamber walls, thus effectively setting $\gamma \equiv \gamma_{\rm so} = \gamma_{\rm ow} = \gamma_{\rm sw}$.

We assume that there are purely repulsive forces between the droplets and hopper walls or obstacles, which are modeled using the following pairwise interaction potential:
\begin{equation}
\label{dp_wall}
U_{w} = \sum_{i=1}^{N_v} \frac{\epsilon_{w}}{2} \left(1-\frac{2d_i}{\delta} \right)^2\Theta \left(1-\frac{2d_i}{\delta}\right),
\end{equation}
where $\epsilon_w$ sets the strength of the repulsive interactions, $d_i$ is the minimum distance between the center of vertex $i$ of the droplet and the wall or obstacle surface, and $\delta$ is the vertex diameter. The Heaviside step function $\Theta(\cdot)$ ensures that the pair forces are non-zero only between vertex $i$ overlaps a chamber wall or obstacle. 

The total potential energy $U$ for the droplet is the sum of the shape-energy function $U_{s}$, gravitational potential energy $U_{g}$, and particle-wall interactions $U_{w}$,
\begin{equation}
\label{totalU}
U =  U_{s} + U_{g} +U_{w},
\end{equation}
where $U_{g} = -M g h$, $h$ is the height of the center of mass of the droplet, $g$ is the gravitational acceleration, $M = \rho a_{0}$ is the mass of the droplet with areal mass density $\rho$ and area  $a_{0} = \pi \sigma^2/4$. 

We also consider dissipative forces from the background fluid acting on each droplet vertex $i$:
\begin{equation}
\vec{F}_{i}^{\zeta} = - \frac{\zeta_i}{N_v} {\vec \nu}_{i},
\end{equation}
where ${\vec \nu}_{i}$ is the velocity of vertex $i$ and $\zeta_i$ is the viscous drag coefficient:
\begin{equation}
\label{eq:drag}
\zeta_i = b_{\infty} \left(1 + \frac{b_0}{b_{\infty}} \left(\frac{\sigma}{d_i}\right)^{\alpha}\right).
\end{equation}
In Eq.~\ref{eq:drag}, $b_{\infty}$ is the drag coefficient when the droplet is far from the walls, $d_i$ is the minimum distance between the center of vertex $i$ and the chamber wall (or obstacle), and $b_0$ is the near-wall drag coefficient. The exponent $\alpha= 4/3$ is calibrated to the experiments on single droplet flows in narrow channels described in Appendix A.

Thus, for the DP model, the equations of motion for the position ${\vec r}_i$ of droplet vertex $i$ is
\begin{equation}
M_{i} \frac{\partial^2 \vec{r}_{i}}{\partial t^2} = -\vec{\nabla}_{r_{i}}U + \vec{F}_{i}^{\zeta},
\end{equation}
where $M_{i} = M/N_v$ is the mass of vertex $i$. 

From Eqs.~\ref{shape_energy}-\ref{dp_wall}, we introduce the dimensionless line tension $\Gamma = \gamma /(g\rho \sigma^2)$ and three additional dimensionless energy scales for the DP model in a gravitational field in 2D: {the dimensionless area compressibility $K_a = k_a\sigma/(g\rho)$, the dimensionless perimeter elasticity $K_l = k_l/(g\rho\sigma)$, and the dimensionless wall interaction energy $E_w = \epsilon_{w}/(g\rho \sigma^3)$.} In these studies, we vary $\Gamma$ in the range $10^{-2} < \Gamma < 10$. We set $K_a>10^4$ so that the fluctuations in the droplet areas are small, $K_l/\Gamma < 10^{-3}$ so that the line tension energy dominates the perimeter elasticity (see Appendix B), and $E_w = 10^4$ to minimize the vertex-wall (and vertex-obstacle) overlaps. The time $t$ is measured in units of $t_0 = \sqrt{\sigma/g}$ and the equations of motion are integrated using a modified velocity-Verlet algorithm with time step $\Delta t=10^{-3} t_0$. 

\subsection{Soft particle model}

We will also describe gravity-driven single-droplet flows through obstacle arrays using the soft particle (SP) model, and compare the results from the SP model to those of the DP model to better understand the influence of droplet shape change on the flows. In general, there are three contributions to the total potential energy: 1) the shape-energy function $U_s$, 2) the gravitational potential energy $U_g$, and 3) the droplet-obstacle interaction energy $U_w$. For the SP model, $U_s=0$ and excluded volume interactions between the droplet and obstacles are generated by allowing overlaps between them\cite{Durian1995,Durian1997,Hong2017,Tao2021}. The pairwise droplet-obstacle interaction energy is
\begin{equation}
\label{sp_wall}
U_{w} =  \frac{\epsilon_{\rm w}}{2} \left(1- \frac{2 d}{\sigma}\right)^2\Theta \left(1- \frac{2 d}{\sigma}\right),
\end{equation}
where $\sigma$ is the SP droplet diameter, $d = r - \sigma_{\rm ob}/2$ is the distance between the center of the droplet and the obstacle surface, $\sigma_{\rm ob}$ is the obstacle diameter, $r$ is the separation between the center of the droplet and the center of the closest obstacle, and $\epsilon_{\rm w}$ is the characteristic energy of the repulsive interactions. The Heaviside step function $\Theta(\cdot)$ ensures that the pair forces are non-zero only when the droplet overlaps with an obstacle. 
Thus, the total potential energy of an SP droplet in an obstacle array is
\begin{equation}
\label{totalU_forSP}
U = U_{g} +U_{w}.
\end{equation}

For the SP model, we consider following form for the viscous drag forces on droplets moving in a background viscous fluid:
\begin{equation}
\vec{F}_{\zeta} = - \zeta_{\rm sp} \vec{v},
\end{equation}
where $\zeta_{\rm sp}$ is the drag coefficient and $\vec{v}$ is the velocity of the SP droplet. We note that, unlike Eq.~\ref{eq:drag} for the DP model, we set $\zeta_{\rm sp} = b_{\infty}$. In contrast to the DP model, the SP model considers droplet deformation by allowing overlaps between the droplets and obstacles, which makes it difficult to describe distance-dependent viscous drag. For the SP model, the equations of motion for the droplet position ${\vec r}_i$ are
\begin{equation}
\label{eom}
M \frac{\partial^2 \vec{r}}{\partial t^2} = -\vec{\nabla}_{r}U + \vec{F}_{\zeta}.
\end{equation}
We integrate Eq.~\ref{eom} using a modified velocity-Verlet integration scheme with time step $\Delta t=10^{-3} t_0$. For the SP model, an important dimensionless energy scale is the ratio of the excluded volume interactions to the gravitational potential energy, $E_{\rm w} = \epsilon_{\rm w}/(g\rho \sigma^{3})$. Here, we vary $E_{\rm w}$ over the range $10^{-1} < E_{\rm w} < 10^2$.

\subsection{Simulation initialization}

For the DP model, we initialize the droplets in narrow channels and obstacle arrays in the shape of regular polygons with edge lengths equal to their equilibrium values $l_{0} = \sqrt{4a_{0} N_v \tan(\pi/N_v)}/N_v$.  For droplet flows in narrow channels, we initially place the droplet far from the orifice, typically at $h = -10 \sigma$, so that the droplet reaches terminal speed before touching the channel walls. For droplets in obstacle arrays, we initially place the droplet in the center of the simulation box. However, this placement typically causes overlaps between the droplet and nearby obstacles. Therefore, after the initial placement of the droplet, we carry out energy minimization (in the absence of gravity) to eliminate overlaps between the droplet and obstacles. We then integrate the equations of motion with gravity and the droplet begins to flow through the obstacle array until it reaches steady state or clogs. (See Sec.~\ref{clog_sec}.)

\section{Experimental Methods}
\label{expmethods}

Below, we will compare the SP and DP simulation results for gravity-driven droplet flows in narrow channels and obstacle arrays to experimental studies of quasi-2D gravity-driven flows of octane oil droplets in water. In this section, we describe the details of our experimental studies. {Note that quantitative experimental studies were performed on flows of capillary droplets through narrow orifices. In contrast, only qualitative experimental studies of droplets flowing through obstacle arrays were performed to illustrate the droplet squeezing and wrapping mechanisms. The experimental studies of droplets in obstacle arrays were only featured in the images in Fig.~\ref{fig:obstaclearray} (a) and (c).}

{For the experiments studying droplet flows through narrow constrictions,} we construct the sample chambers using laser-cut plastic film with thickness $400$~$\mu$m, sandwiched between two glass microscope slides.  The film is attached to the slides using Norland Optical Adhesive~\cite{Yuxuan}.  {The film is cut into two triangular shapes and placed such that they form a narrow orifice, as shown in Fig.~\ref{fig:model}(a).  The slope of the walls coming into the orifice is $45^\circ$.}   The diameters of the droplets are tuned so that they are larger than the chamber thickness to ensure that each droplet is quasi-two-dimensional when confined by the microscope slides.

{We also conduct qualitative experiments of droplets flowing through obstacle arrays to illustrate the physical mechanisms of droplet squeezing and wrapping.  For these qualitative experiments, the obstacles } are produced by placing small drops of UV adhesive between the slides, which then cure into solid cylinders spanning the sample chamber thickness. {In addition, we do not observe any influence of gravity in how the obstacles cure. The chamber is sufficiently thin such that surface tension determines the obstacle shapes. The maximum diameter at the obstacle center is no more than $10\%$ larger than the diameter at the glass walls.}

{Our experiments focus on individual oil droplets passing through the microfluidic chamber, and thus we do not need to produce a large number of droplets.  We form  single oil droplets by direct injection of the oil into the chamber via a hand-held syringe.  The droplet diameter is subsequently measured {\it in situ}.  Different experiments typically use different chambers with their own oil droplet, allowing us to vary the orifice width $w$ and droplet diameter $\sigma$.}  We use octane with density $\rho_{o} = 0.703$ g/ml suspended in water with density $\rho_{w} = 0.997$ g/ml. Despite the absence of other oil droplets, we find that it is still necessary to use a 0.5\% Tween 20 nonionic detergent solution to prevent sticking of the oil droplets to the microscope slide surfaces. 

We image the droplets using a $1.6\times$ lens (0.05 Numerical Aperture) attached to a Leica DM4500 B inverted microscope. Movies are acquired by attaching a $0.35\times$ C-mount to a ThorLabs DCC1545M camera, and recording at a frame rate commensurate with the droplet speed. The microscope is tilted slightly from the horizontal to cause gravitationally driven motion of the droplet.  In particular, the tilt angle of the microscope is changed by placing a spacer underneath the front edge of the microscope, and the angle is measured relative to the ground via a Wixley digital angle gauge.  {As the droplet density is less than water, the droplets rise in the sample chamber due to the buoyant force.  In all images presented in this article, the direction of droplet motion is in the direction of the arrow labeled by $f$.}

The droplet motion is affected by viscous drag from the water phase (viscosity $\eta_{w} \approx 1$~mPa$\cdot$s) and viscous drag between the oil and glass plates ($\eta_{o} \approx 100$~mPa$\cdot$s).  The droplets try to remain round due to surface tension ($\gamma_{\rm ow} \approx 10$~mJ/m$^2$, as measured by the ``Dropometer" from Droplet Labs).  However, under gravity, the droplets deform when pressed against the sample chamber walls. {We do not observe droplet break-up in these experiments.}

{We incorporate quasi-2D effects in the simulations by defining an effective 2D surface tension based on the 3D surface tension $\gamma_{\rm exp}$ from the experiments. The droplet surface energy can be split into two parts, $ E=E^\prime+E_{\rm oop}$, where $E^\prime=\gamma_{\rm exp}(a_{\rm top}+a_{\rm bottom})$  is the surface energy of the top and bottom surfaces of the droplet, and $E_{oop}=\gamma_{\rm exp} a_{\rm oop}$ is the surface energy of the out-of-plane surface of the droplet with area $a_{\rm oop}$ as shown in Fig.~\ref{fig:model} (d). In the quasi-2D limit, the droplet thickness satisfies $H \ll P$, where  $P$ is the perimeter of the droplet from the top view and  $a_{\rm oop} \approx P H$.  Note that $a_{\rm top}+a_{\rm bottom}$ is roughly constant due to fluid incompressibility. Thus, when the droplet deforms, the change in surface energy $\delta E=\delta E^\prime+\delta E_{\rm oop} \approx\gamma_{\rm exp} \delta a_{\rm oop} \approx \gamma_{\rm exp} H \delta P$,  and the effective 2D surface tension is $\gamma=\gamma_{\rm exp} H$. We nondimensionalize $\gamma$ by the buoyancy force such that the dimensionless surface tension is defined as:}
\begin{equation}
\label{eq:expGamma}
\Gamma_{\rm{exp}}=\frac{\gamma}{g_{\rm{exp}}\rho_o\sigma^2H}=\frac{\gamma_{\rm{exp}}}{g_{\rm{exp}}\rho_o\sigma^2}.
\end{equation}
Here, $\gamma_{\rm exp}$ is the oil-water interfacial surface tension, $g_{\rm exp} = \sin\theta g (\rho_w - \rho_o)/\rho_o$ is the acceleration imposed by oil-in-water buoyancy, $g \sim 9.8$ m/s$^2$ is the gravitational constant, $\sigma$ is the average undeformed diameter of the droplets, and $\theta$ is the microscope tilt angle. 

\section{Results}
\label{results}

\begin{figure*}[!th]
\centering
\includegraphics[width=0.85\textwidth]{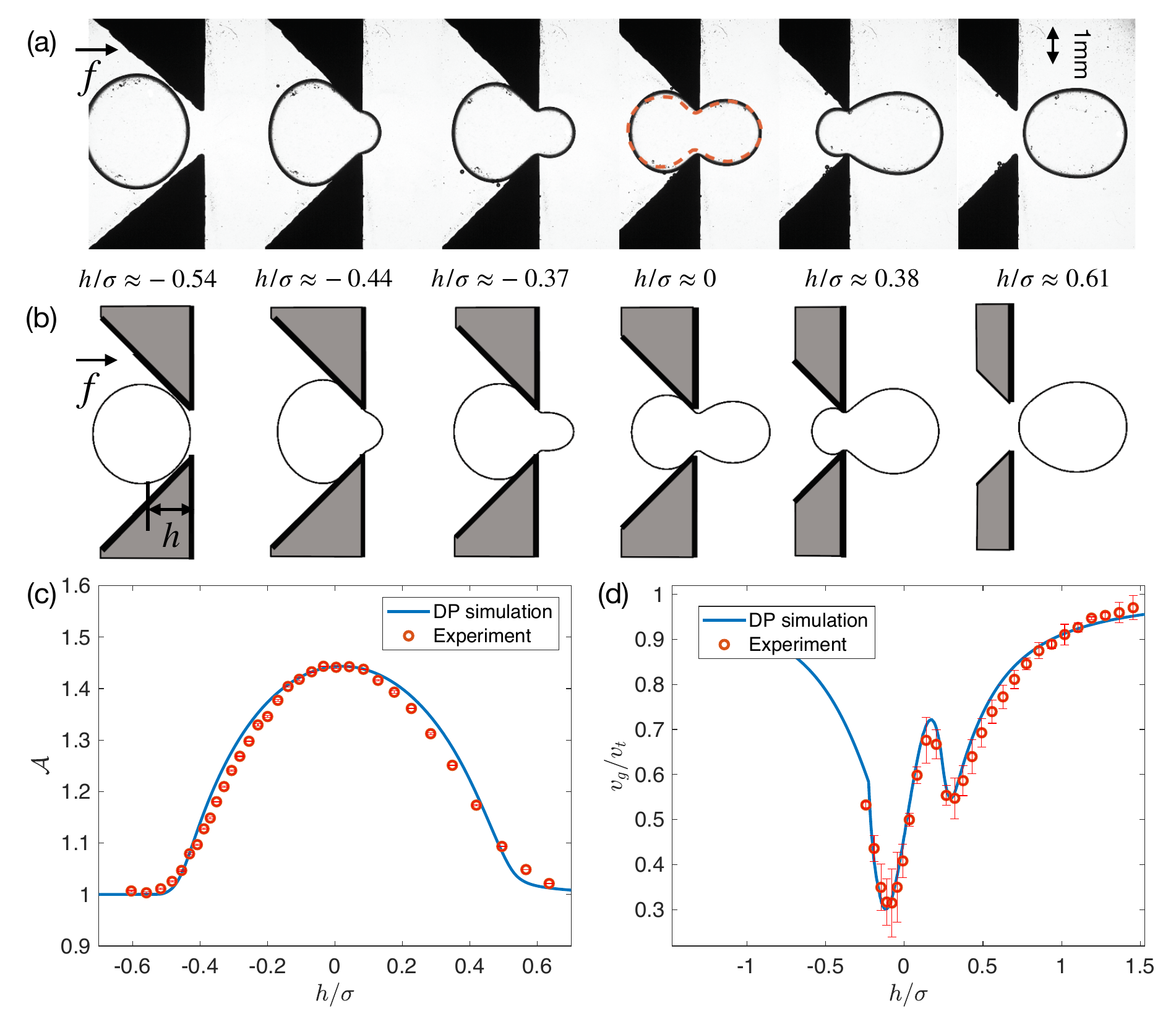}
\centering
\caption{Series of images of a single droplet flowing through a narrow orifice with width $w= 0.4 \sigma$ from (a) oil-in-water experiments with undeformed droplet diameter $\sigma \approx 3.5 ~\rm mm$ and tilt angle $\theta \approx 28^{\circ}$ and (b) a DP simulation in 2D at dimensionless line tension $\Gamma = \Gamma^*$ and near-wall drag coefficient $b_0=b_0^*$. The scale bar indicates $1$~mm. The rightward pointing arrow indicates the direction of {droplet flow}. Below panel (a), we provide the distances of the droplet center of mass to the orifice $h$ at which the images are captured. We find that $\Gamma^* \approx 0.16$ and $b_0^*/b_{\infty} \approx 0.064$ minimize the deviation in the droplet's center of mass speed $\Delta_{v}$ between the DP simulations and experiments. These best-fit values give $\Delta_{v} \approx 0.09$ and $\Delta_{\cal{A}} \approx 0.01$. In panel (a), we overlay the shape of the droplet from the DP simulations at $h=0$ (red dashed line) onto the corresponding droplet image for the experiments (black solid line). Droplet (c) shape parameter $\mathcal{A}$ and (d) center of mass speed in the driving direction $v_g$ plotted as a function of $h/\sigma$ for both experiments (open circles) and DP simulations (solid lines). We estimate the dimensionless surface tension in the experiments to be $\Gamma_{\rm exp} \approx 0.57$.  {The error bars for the experimental data are obtained using the standard deviation of the measured quantities from at least five different trials with one droplet.}}
\label{fig:single}
\end{figure*}

In this section, we describe the results from the numerical simulations of gravitationally-driven droplet flows through narrow channels and obstacle arrays. We first present a method to calibrate the DP simulations to experimental results for droplet flows through narrow channels. We then show results for the droplet speed in the direction of the driving $\nu_g$ as a function of distance from the orifice. We find that the droplet speed profile is nonmonotonic and can even possess an overshoot, such that the droplet speed exceeds the terminal speed $\nu_t$ as it exits the orifice.  Using the DP and SP models, we also investigate gravity-driven droplet flows through obstacle arrays.  For the DP simulations, we find non-monotonic behavior for the clogging probability as a function of the dimensionless line tension $\Gamma$.  We show that droplets can clog at small $\Gamma$ when they are highly deformable and can wrap around the obstacles. We emphasize that the SP model does not include the wrapping mechanism since wrapping requires significant changes in the droplet shape. As shown in many previous studies of flows of nearly hard particles in confined geometries, the droplets can also clog when they are nearly rigid at large $\Gamma$. At intermediate $\Gamma$, we find a regime of continuous droplet flows through the obstacle arrays.  We observe qualitatively similar results over a wide range of obstacle densities.

\begin{figure*}[!ht]
\centering
\includegraphics[width=0.9\textwidth]{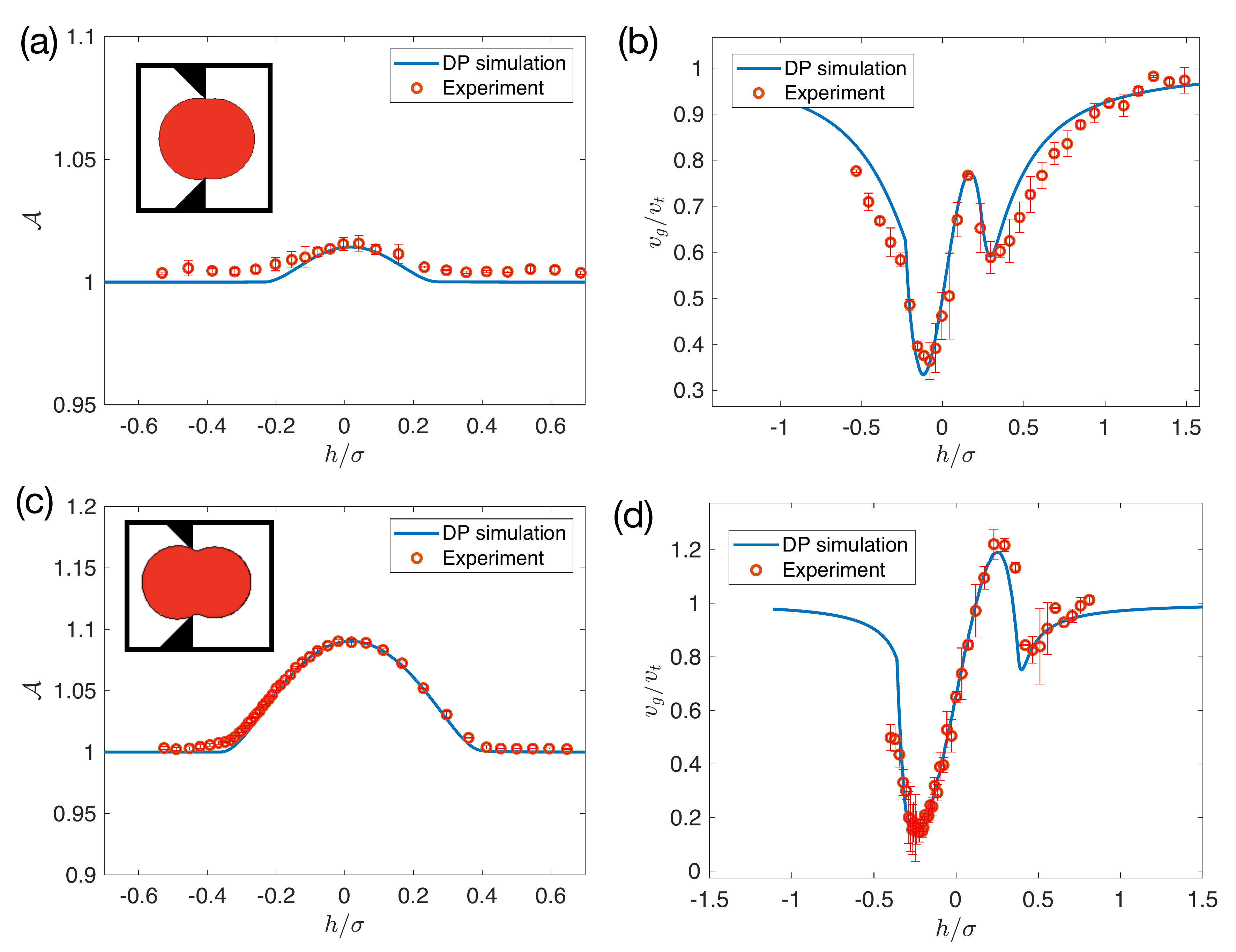}
\centering
\caption{Droplet (a) shape parameter $\mathcal{A}$ and (b) center of mass speed $v_g$ plotted as a function of distance from the orifice $h/\sigma$ for both experiments (open circles) and DP simulations (solid lines) at $w =0.9 \sigma$. For (a) and (b), the dimensionless surface tension $\Gamma_{\rm exp} \approx 5.80$, tilt angle $\theta \approx 8.5^{\circ}$, and undeformed droplet diameter $\sigma \approx 1.98$~mm. After minimizing $\Delta_{v}$ between the experiments and DP simulations, we obtain $\Gamma^* \approx 1.7$, $b_0^*/b_{\infty} \approx 0.21$, $\Delta_{v} \approx 0.08$, and $\Delta_{\cal{A}} \approx 0.004$.  The inset to (a) shows the droplet shape from the DP simulations at maximum deformation with $\mathcal{A}_{\rm max} \approx 1.014$. (c) $\mathcal{A}$ and (d) $v_g$ plotted as a function of $h/\sigma$ for both experiments and DP simulations at $w =0.7 \sigma$. For (c) and (d), $\Gamma_{\rm exp} \approx 3.47$, $\theta \approx 5.0^{\circ}$, and $\sigma \approx 3.30$~mm. After minimizing $\Delta_{v}$, we obtain $\Gamma^* \approx 1.08$, $b_0^*/b_{\infty} \approx 0.07$, $\Delta_{v} \approx 0.06$, and $\Delta_{\cal{A}} \approx 0.005$.  The inset to (c) shows the droplet shape from the DP simulations at maximum deformation with $\mathcal{A}_{\rm max} \approx 1.093$.  {The error bars for the experimental data are obtained using the standard deviation of the measured quantities from at least five different trials with one droplet.}}
\label{fig:vprofile}
\end{figure*}

\subsection{Calibration of DP model using experiments of a single droplet flowing through narrow channels}
\label{calibrated}

We first carried out both experiments and DP simulations of a single droplet flowing through narrow channels under the influence of gravity. To enable a comparison between the experimental and simulation results, we define the root-mean-square deviations in the droplet shape parameter $\Delta_{\cal{A}}$ and speed $\Delta_{\nu}$:
\begin{equation}
\label{deltaA}
   \Delta_{\cal{A}} = \sqrt{\sum_{i=1}^{N_f} \frac{(({\cal{A}}_{\rm exp,i}-{\cal{A}}_{\rm sim,i})/{{\cal{A}}_{\rm sim,i}})^2}{N_f}}
\end{equation}
and
\begin{equation}
\label{deltav}
    \Delta_{\nu} = \sqrt{\sum_{i=1}^{N_f} \frac{((\nu_{\rm exp,i} - \nu_{\rm sim,i})/\nu_{\rm sim,i})^2}{N_f}},
\end{equation}
where ${\cal{A}}_{\rm exp,i}$ is the droplet shape parameter in the $i$th frame in the experimental video, and ${\cal{A}}_{\rm sim,i}$ is the droplet shape parameter in the DP simulations when the droplet center of mass is at the same location as it is in the experiments at frame $i$. Similarly, $\nu_{\rm exp,i}$ and $\nu_{\rm sim,i}$ are the droplet speed in the direction of gravity at the $i$th frame in the experiments and simulations. The summations in Eqs.~\ref{deltaA} and~\ref{deltav} are over all $N_f$ frames in the experimental videos. Despite using a small time step in the DP simulations, we might not find a frame in the simulations where the droplet has the same position as that in the experiments. In this case, we use linear interpolation between frames to generate ${\cal{A}}_{\rm sim,i}$ and $\nu_{\rm sim,i}$.

We seek to minimize $\Delta_{\cal{A}}$ and $\Delta_\nu$ by tuning the dimensionless line tension $\Gamma$ and near-wall drag coefficient $b_0/b_{\infty}$ in the DP simulations. In practice, $\Delta_{\cal{A}}$ is more than an order of magnitude smaller than $\Delta_\nu$, and thus we only minimize $\Delta_v$ over $\Gamma$ and $b_0/b_{\infty}$.  $b_{\infty}$ is determined by matching the droplet's terminal speed far from wall in the DP simulations and experiments. In particular, we equate the dimensionless terminal speed, 
\begin{equation}
\label{eq:bInfty}
    V_t \equiv \frac{v_t}{\sqrt{\sigma g_{\rm exp}}}=\frac{m g }{b_\infty\sqrt{\sigma g}},
\end{equation}
in the experiments and DP simulations.

In Fig.~\ref{fig:single} (a) and (b), we show the droplet shapes as a function of distance $h$ from the orifice (with width $w = 0.4 \sigma$) from trajectories of gravity-driven droplet flows through narrow channels in experiments and DP simulations using the values $\Gamma^* = 0.16 \pm 0.01$ and $b_0^*/b_{\infty} = 0.064 \pm 0.003$ that minimize $\Delta_{\nu}$.  The error bars for $\Gamma^*$ and $b_0^*$ are the standard deviations in $\Gamma^*$ and $b_0^*$ from fitting the the DP simulations to at least five independent experimental trials. For the experiments in Fig.~\ref{fig:single} (a), the terminal velocity $v_t \approx 7$~mm/s far from the wall, undeformed droplet diameter $\sigma \approx 3.5 ~\rm mm$, and tilt angle $\theta \approx 28^{\circ}$. Thus, for the data in Fig.~\ref{fig:single} (a), we estimate the dimensionless surface tension for the droplets to be $\Gamma_{\rm exp} \approx 0.57$ and the dimensionless terminal velocity to be $V_t \approx 0.08$. In Fig.~\ref{fig:single} (c) and (d), we plot $\mathcal{A}$ and $v_g$ as a function of $h/\sigma$ for both the calibrated DP simulations and experiments for the data presented in Fig.~\ref{fig:single} (a) and (b). We find that the droplet shapes and speed profiles for the DP simulations and experiments are similar over the full range of $h$, with $\Delta_{\cal{A}} \approx 0.01$ and $\Delta_{v} \approx 0.09$. Note that despite the simplicity of the 2D DP simulations, $\Gamma^*$ is comparable to $\Gamma_{\rm exp}$ (with $\Gamma^*/\Gamma_{\rm exp} \approx 0.3$), emphasizing the predictive power of 2D DP simulations. {We also note that, in general the shape parameter is not symmetric about $h=0$. For example, in Fig.~\ref{fig:single} (c) the shape parameter at $h\approx 0.5\sigma$ is $\mathcal{A}\approx1.07$, whereas at $h\approx-0.5\sigma$, it is $\mathcal{A}\approx1.00$.}

In Fig.~\ref{fig:vprofile}, we compare the results for the calibrated DP simulations and experiments of gravity-driven droplet flows in narrow channels for two additional experiments with different droplet sizes, orifice widths and tilt angles. In Fig.~\ref{fig:vprofile} (a) and (b), we show ${\cal A}$ and $v_g$ versus $h/\sigma$ for $w/\sigma = 0.90$, $\theta \approx 8.5^{\circ}$, $\sigma \approx 1.98$~mm, $v_t \approx 0.97$~mm/s, $V_t \approx 0.028$, and $\Gamma_{\rm exp} \approx 5.80$. In Fig.~\ref{fig:vprofile} (c) and (d), we show ${\cal A}$ and $v_g$ versus $h/\sigma$ for $w/\sigma=0.70$, $\theta \approx 5.0^{\circ}$, $\sigma \approx 3.30$~mm, $v_t \approx 1.1$~mm/s, $V_t \approx 0.032$, and $\Gamma_{\rm exp} \approx 3.47$. 
{
In the DP simulations we can obtain the dimensionless line tension $\Gamma^{**}$ 
for droplets in the two additional experiments that minimizes $\Delta_{\nu}$ for the experimental data in Fig.~\ref{fig:vprofile} using $\Gamma^{**}/\Gamma^* = (\sigma^*)^2 \sin\theta^*/((\sigma^{**})^2 \sin\theta^{**})$, where $\theta^*$, $\sigma^*$ and $\Gamma^*$ are the tilt angle, undeformed droplet diameter and fitted dimensionless line tension for the data in Fig.~\ref{fig:single}, and $\theta^{**}$, $\sigma^{**}$ are the tilt angle and undeformed droplet diameter for the two additional experiments in Fig.~\ref{fig:vprofile}.} We find $\Gamma^{**} = 1.7 \pm 0.1$ for $\theta^{**} = 8.5^{\circ}$ and $\sigma^{**} \approx 1.98$~mm (with $\Gamma_{\rm exp} \approx 5.8$) and $\Gamma^{**} = 1.08 \pm 0.09$ for $\theta^{**} =5.0^{\circ}$ and $\sigma^{**} = 3.30$~mm (with $\Gamma_{\rm exp} \approx 3.47$).  For all tilt angles and droplet diameters studied, we find that the ratio of the dimensionless line tension that minimizes $\Delta_v$ to the dimensionless surface tension in experiments satisfies $\Gamma^{**}/\Gamma_{\rm exp} \approx 0.3$, which likely stems from the fact that we compare values of the dimensionless line tension in 2D to values of the dimensionless surface tension in 3D. We find that the near-wall drag coefficients that minimize $\Delta_v$ do not depend strongly on droplet diameter and tilt angle, $b_0^*/b_{\infty} \approx 0.06$-$0.2$.

\begin{figure}[bth]
\centering
\includegraphics[width=0.95\columnwidth]{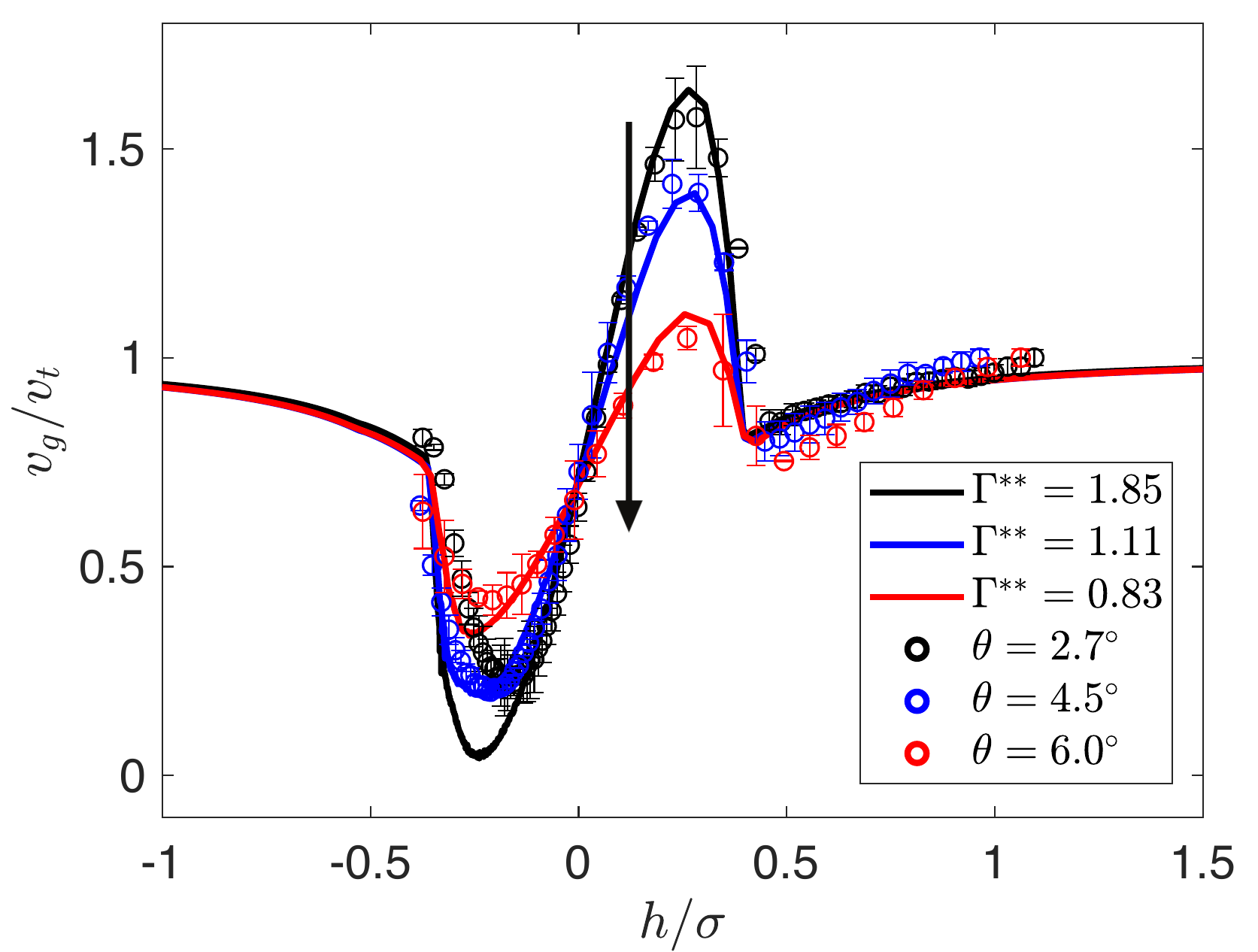}
\centering
\caption{Droplet center of mass speed in the direction of the gravitational driving (normalized by the terminal speed) $v_g/v_t$ plotted as a function of $h/\sigma$ for droplet flows in narrow channels with orifice width $w=0.7\sigma$ in the experiments (open circles) for several values of the tilt angle, $2.7^\circ$ (black), $4.5^\circ$ (blue), and $6.0^\circ$ (red), and calibrated DP simulations (solid lines) with fitted line tensions $\Gamma^{**} = 1.85$ (black), $1.11$ (blue), and $0.83$ (red). The downward arrow indicates increasing tilt angle $\theta$.{The error bars for the experimental data are obtained using the standard deviation of the measured quantities from at least five different trials with one droplet.}}
\label{fig:overshoot}
\end{figure}

A key feature of the droplet speed profile $v_g$ is that it can possess nonmonotonic behavior and $v_g$ can even exceed the terminal speed $v_t$ as the droplet exits the orifice. For example, in Fig.~\ref{fig:vprofile} (d), $\nu_g$ overshoots $v_t$ near $h/\sigma \approx 0.3$. Similar phenomena have been observed in pressure-driven droplet flows in obstacle arrays, where the pressure drop across the orifice changes from positive to negative, which indicates that the droplet speed exceeds the background fluid velocity\cite{Coelho2023}. To further investigate the effects of the gravitational driving on the nonmonotonic speed profile, we show $\nu_g$ versus $h/\sigma$ for fixed $w/\sigma=0.7$ for several values of the tilt angle: $\theta=2.7^\circ$, $4.5^\circ$, and $6^\circ$ in Fig.~\ref{fig:overshoot}. The nonmonotonic droplet speed profile can be understood by analyzing energy transfer while the droplet moves through the narrow channel. The droplet first impacts the chamber wall at $h/\sigma < 0$ and deforms while squeezing through narrow orifice. During this process, the kinetic and gravitational energy of the droplet transfers into surface energy of the deformed droplet, which corresponds to the first speed minimum near $h/\sigma \approx -0.25$. The large surface energy is transferred into kinetic energy as the droplet continues to move through the orifice, which can give rise to speeds that exceed the terminal velocity near $h/\sigma \approx 0.3$. However, the excess kinetic energy is quickly dissipated by viscous drag, and $\nu_g$ decreases again, reaching a second local minimum near $h/\sigma \approx 0.5$. Finally, the speed increases again, approaching $\nu_t$ as the droplet moves far from the orifice since the drag coefficient $\zeta$ decreases toward $b_{\infty}$ for $h/\sigma \gg 1$. (See Eq.~\ref{eq:drag}.)

As shown in Fig.~\ref{fig:overshoot}, the amplitude of the overshoot in the droplet speed increases with decreasing $g$ and $\theta$. The excess shape energy $\Delta U_s$ gained during droplet deformation is largely determined by the width of the orifice, $w/\sigma$. In contrast, the droplet kinetic energy at the terminal speed $E_t$ decreases with increasing $g$. Thus, for $h > 0$, the excess surface energy $\Delta U_s$ transferred into kinetic energy $E_k \sim \Delta U_S$, generates enhanced droplet speed overshoots $\nu_g \sim \sqrt{E_k/E_t}$ as $g$ or $\theta$ decreases. {These results are consistent with the intuition that, the dimensionless surface tension $\Gamma_{\rm exp}$ in Eq.~\ref{eq:expGamma} is inversely proportional to the gravitational acceleration g, which in turn is proportional to the tilt angle. Thus, $\Gamma_{\rm exp}$ is inversely proportional to the tilt angle. As a result, increasing the tilt angle, decreases $\Gamma_{\rm exp}$, which in turn decreases the optimal fit $\Gamma^{**}$ and reduces the overshoot.}

\subsection{Clogging of single droplets in obstacle arrays}
\label{clog_sec}

\begin{figure}[h]
\centering
\includegraphics[width=0.9\columnwidth]{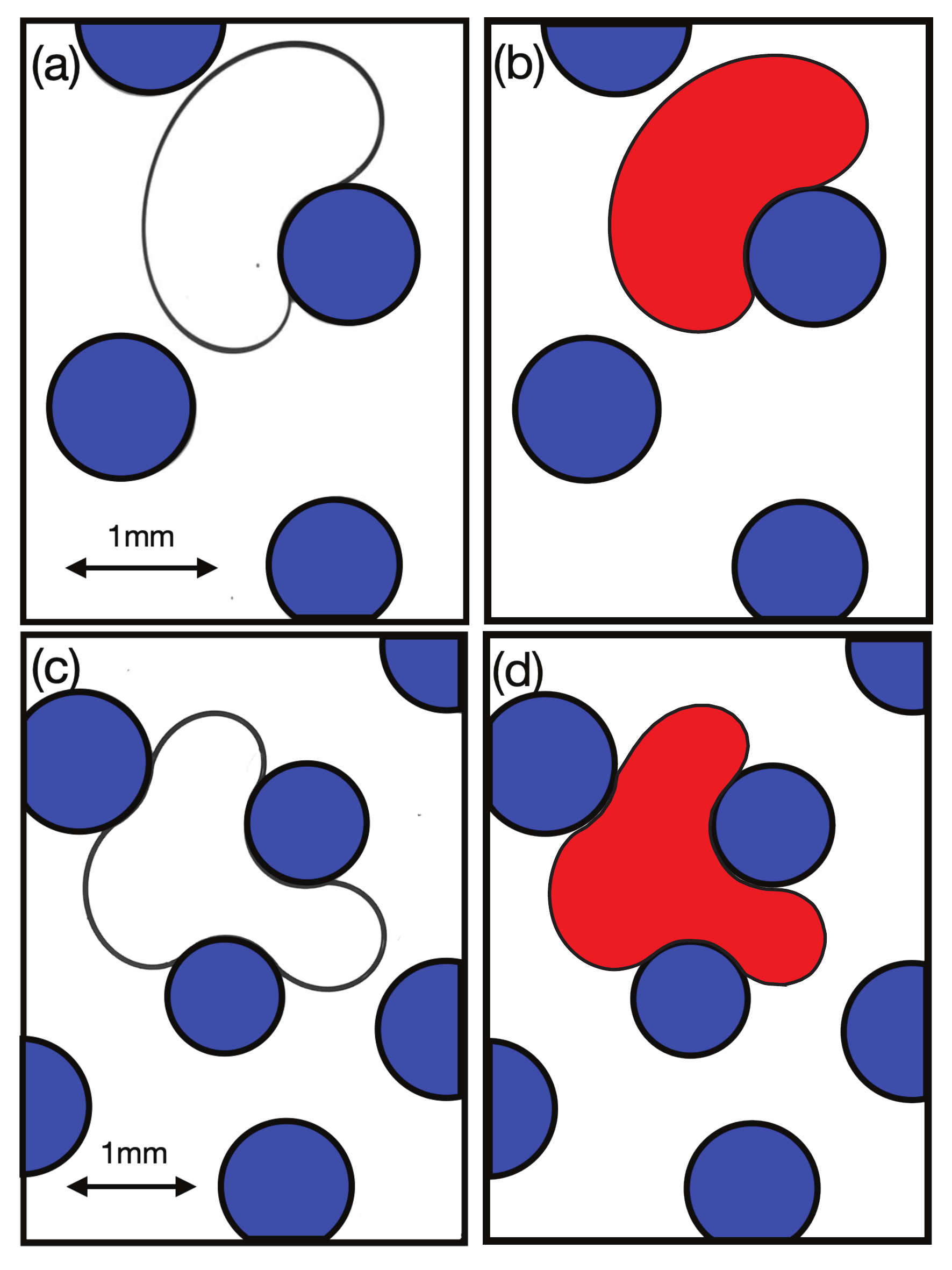}
\centering
\caption{Oil-in-water droplets flowing through an obstacle array can wrap around obstacles in (a) experiments with $\Gamma_{\rm exp} \approx 0.88$, tilt angle $\theta \approx 90^{\circ}$, undeformed droplet diameter $\sigma \approx 2$~mm, obstacle diameter $\sigma_{\rm ob}/\sigma \approx 0.3$-$0.4$, and minimum obstacle separation $w_{\rm ob}/\sigma \approx 0.2$ and (b) calibrated DP simulations with $\Gamma \approx 0.3$ and a similar obstacle array to that used in (a). In experiments, the driving force is directed from left to right and in the DP simulations the flow direction is to the right. The droplets can also squeeze through narrow constrictions formed by adjacent obstacles in a similar set of (c) experiments and (d) DP simulations as those in (a) and (b).}
\label{fig:obstaclearray}
\end{figure}

Droplet shape dynamics plays an important role in determining the motion and clogging of droplets flowing through obstacle arrays. For example, in Fig.~\ref{fig:obstaclearray} (a) and (c), we show that droplets can wrap around obstacles and squeeze through narrow constrictions formed by adjacent obstacles. These shape change mechanisms do not occur for nearly rigid particles and can be accurately modeled using the calibrated DP simulations with surface tension described in Sec.~\ref{calibrated}, as shown in Fig.~\ref{fig:obstaclearray} (b) and (d).

For minimum obstacle separation $w_{\rm ob}/\sigma \gg 1$, single droplets flow continuously through obstacle arrays without clogging. However, when $w_{\rm ob}/\sigma <1$, single droplets can become permanently clogged within the obstacle array. We assume that as the droplet moves through the obstacle array, it successively interacts with each obstacle, with each interaction contributing to the  probability of clogging.  This scenario suggests that particle clogging is a Poisson process, confirmed by numerous prior studies\cite{hong22,PRLThomas,Marin2019}.  This means that the droplet can move a distance $r>0$ without clogging with cumulative probability distribution function, 
\begin{equation}
P(r) =  e^{-r/\lambda},
\label{eq:decaylength}
\end{equation}
where $\lambda$ is clogging decay length such that $ P(\lambda)=e^{-1}$.  {We define clogging in the simulations to occur when the droplet’s dimensionless kinetic energy falls below a small threshold. Additional details concerning clogging and the} measurements of $P(r)$ in the DP simulations are included in Appendix C.

\begin{figure}[h]
\centering
\includegraphics[width=0.95\columnwidth]{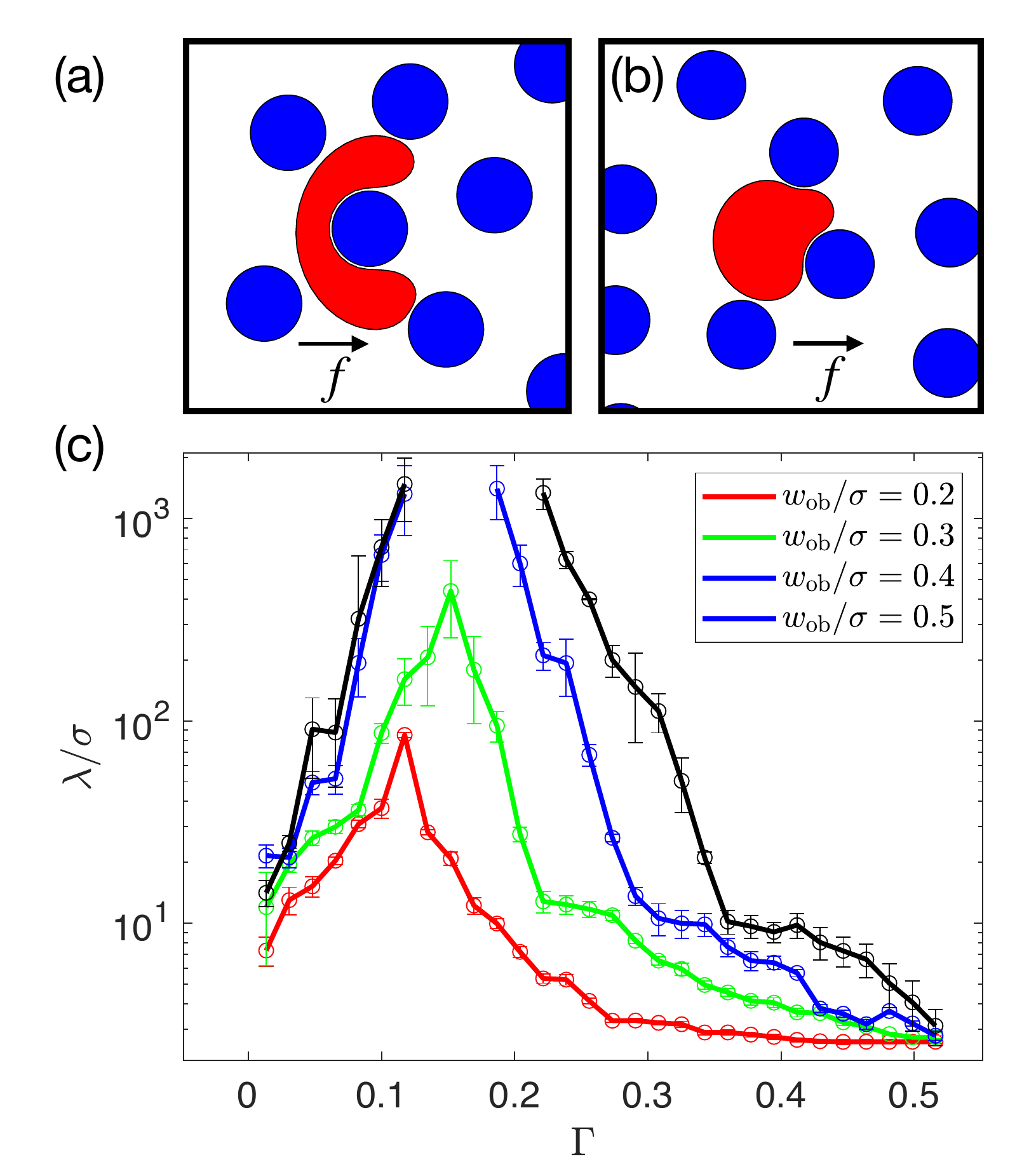}
\centering
\caption{Images of a single droplet (shaded red) from DP simulations moving through arrays of obstacles (shaded blue) with random positions. The rightward pointing arrow indicates the direction of {droplet flow}. Clogs form via the (a) wrapping and (b) squeezing mechanisms. (c) Clogging decay length $\lambda/\sigma$ plotted as a function of line tension $\Gamma$ for the obstacle arrays in (a) and (b) with $\sigma_{\rm ob}/\sigma = 0.3$ and $w_{\rm ob}/\sigma=0.2$ (red), $0.3$ (green), $0.4$ (blue), and $0.5$ (black).}
\label{fig:clogp}
\end{figure}

Several previous studies of clogging of soft particles in hoppers have shown that softer particles (i.e. with smaller elastic moduli) are less likely to form clogs~\cite{Hong2017,Tao2021,D2SM00318J}, and thus would possess larger values of the clogging decay length $\lambda$. Does this result also hold for the highly deformable droplets considered in this study? To address this question, we carry out numerical simulations of single droplets (using the SP and DP models) moving through obstacle arrays in the clogging regime with $w_{\rm ob}/\sigma \leq 0.5$. We measure $\lambda$ as a function of $\Gamma$ (for the DP model), $K_{\rm w}$ (for the SP model), and the deformability $\delta_s/\sigma$, given by the fraction of the droplet diameter that deforms under its own weight. We find that the wrapping mechanism plays an important role in determining the clogging statistics for single droplets moving through obstacle arrays. In particular, clogs can form when droplets {\it wrap} around an obstacle as shown in Fig.~\ref{fig:clogp} (a). The droplet is mainly supported by the obstacle around which the droplet has wrapped, with additional stabilizing forces from adjacent obstacles. For less deformable particles, clogs form when droplets cannot {\it squeeze} through narrow channels between adjacent obstacles as shown in  Fig.~\ref{fig:clogp} (b).

In Fig.~\ref{fig:clogp} (c), we plot $\lambda/\sigma$ as a function of the line tension $\Gamma$ from DP model simulations of single droplets moving through obstacle arrays for $\sigma_{\rm ob}/\sigma = 0.3$ and $w_{\rm ob}/\sigma = 0.2$, $0.3$, $0.4$, and $0.5$. We find that $\lambda/\sigma$ versus $\Gamma$ is non-monotonic. At large $\Gamma$, where droplet shape deformation is small, decreasing $\Gamma$ increases $\lambda$, which implies that softer particles are less likely to clog. However, at small $\Gamma$, where droplet shape deformation is large, we find the opposite behavior.  In this regime, decreasing $\Gamma$ {\it decreases} $\lambda$, which implies that softer particles are more likely to clog. The nonmonotonic behavior of $\lambda(\Gamma)$ is caused by the two different mechanisms for clogging (wrapping versus squeezing) as shown in Fig.~\ref{fig:clogp} (a) and (b). At large $\Gamma$, droplets become clogged when squeezing between obstacles. It is easier for nearly rigid droplets to squeeze through the gaps between obstacles as $\Gamma$ decreases, which gives rise to decreased clogging probability and increased $\lambda$. In contrast, at small $\Gamma$, highly deformable droplets can easily squeeze through narrow gaps between obstacles, but they can wrap around an obstacle and clog. Thus, increased deformability enhances the wrapping mechanism, increasing the clogging probability, and decreasing $\lambda$. In Fig.~\ref{fig:clogp} (c), we also show that $\lambda$ increases with $w_{\rm ob}$ as expected, since wider gaps between obstacles reduce clogging. {We also find that the transition from wrapping to squeezing behavior shifts to larger values of $\Gamma$ as $w_{\rm ob}/\sigma$ increases. This shift occurs because wrapping is primarily a single-obstacle phenomenon. Larger separation between obstacles makes wrapping more prevalent, while narrower gaps between obstacles promotes squeezing behavior.}

\begin{figure}[!h]
\centering
\includegraphics[width=0.95\columnwidth]{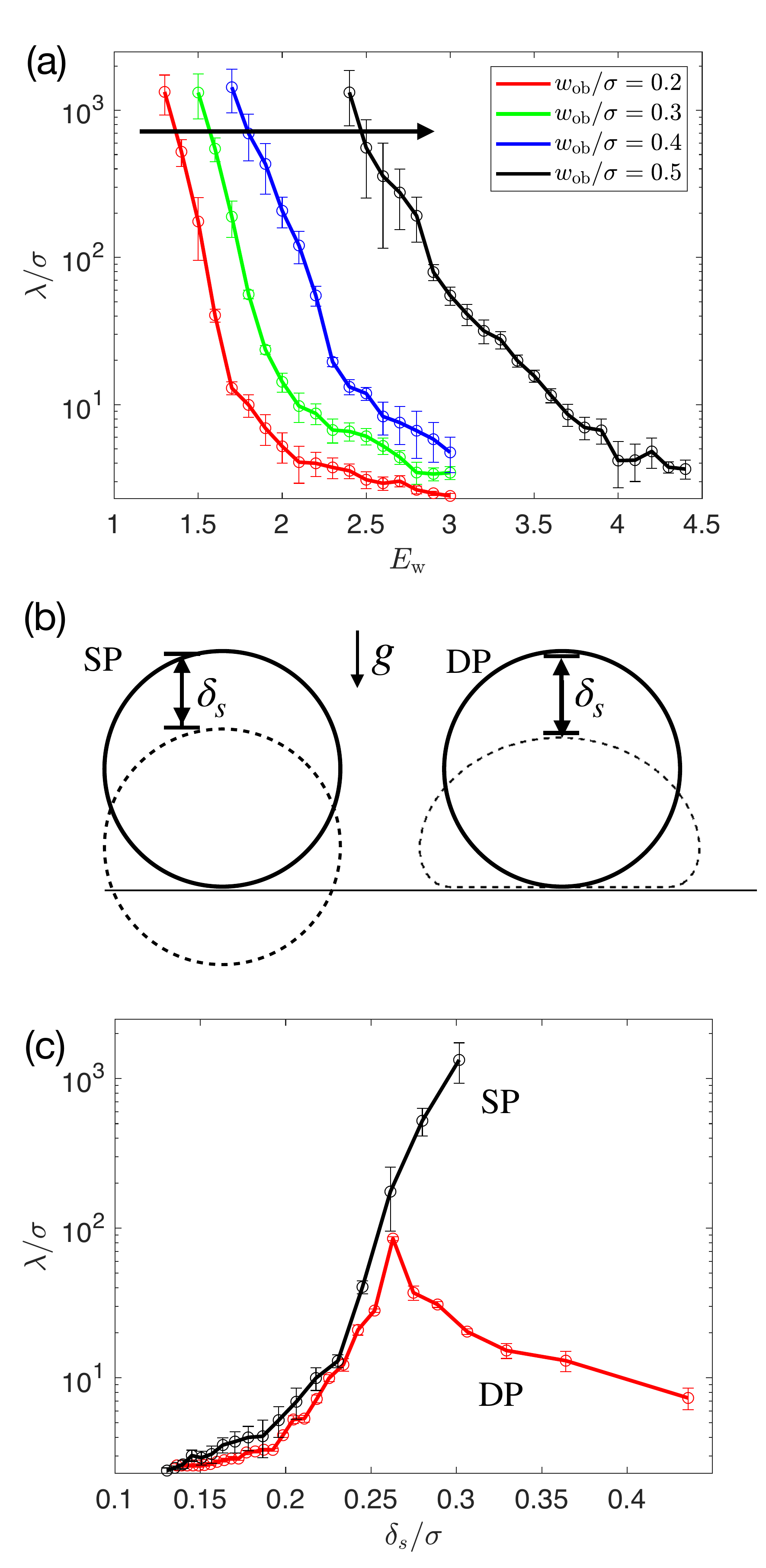}
\centering
\caption{(a) Clogging decay length $\lambda/\sigma$ plotted as a function of normalized stiffness $E_{\rm w}$ from SP model simulations of single droplets moving through obstacle arrays with $\sigma_{\rm ob}/\sigma = 0.3$ and $w_{\rm ob}/\sigma=0.2$ (red), $0.3$ (green), $0.4$ (blue), and $0.5$ (black). The arrow indicates increasing $w_{\rm ob}/\sigma$. (b) Schematic of the deformation $\delta_s$ that a droplet undergoes under its own weight for the (left) SP and (right) DP models. The solid (dashed) outlines of the droplet represent the droplet before (after) gravity is added. (c) $\lambda/\sigma$ plotted as a function of $\delta_s/\sigma$ for both SP (black) and DP (red) models for obstacle arrays with $w_{\rm ob}/\sigma=0.2$.}
\label{fig:SP_p}
\end{figure}

We also carry out SP simulations of single droplets moving through the same obstacle arrays in Fig.~\ref{fig:clogp} to understand the effects of droplet stiffness versus deformability on clogging. In Fig.~\ref{fig:SP_p} (a), we plot $\lambda/\sigma$ versus the dimensionless stiffness $E_{\rm w}$ for the SP model; $\lambda/\sigma$ decreases {\it monotonically} with increasing $E_{\rm w}$ over the full range of $E_{\rm w}$, and as expected, $\lambda/\sigma$ increases with $w_{\rm obs}/\sigma$. These results emphasize that the SP model can only generate clogs via the squeezing mechanism, not the wrapping mechanism.

To directly compare the results for single droplet motion through obstacle arrays using the SP and DP models, we consider the droplet deformation $\delta_s$ caused by its own weight~\cite{Hong2017}. (See Fig.~\ref{fig:SP_p} (b).) To calculate $\delta_s$, we initially turn off gravity and place an SP droplet or a DP droplet with diameter $\sigma$ next to a wall. We then turn on gravity and measure the distance $\delta_s$ that the droplet sinks. In Fig.~\ref{fig:SP_p} (c), we plot $\lambda/\sigma$ versus $\delta_s/\sigma$ for both the SP and DP models of single droplet motion through obstacle arrays. $\lambda/\sigma$ versus $\delta_s$ is similar for the SP and DP models for $\delta_s/\sigma < 0.25$, but the results differ significantly for $\delta_s/\sigma > 0.25$. At large $\delta_s/\sigma$, overlaps between the obstacles and SP droplets are not strongly penalized, clogs are highly unlikely, and $\lambda/\sigma \rightarrow \infty$. However, at large $\delta_s/\sigma$, DP droplets are extremely deformable, can wrap around obstacles, and thus $\lambda/\sigma \rightarrow 0$. Comparing the results for the SP and DP models allows us to assess the importance of shape deformability, particle softness, and volume conservation in determining the clogging probability. 

\subsection{Continuous single-droplet flows in obstacle arrays}
\label{sec:v_g}

\begin{figure}[h]
\centering
\includegraphics[width=0.95\columnwidth]{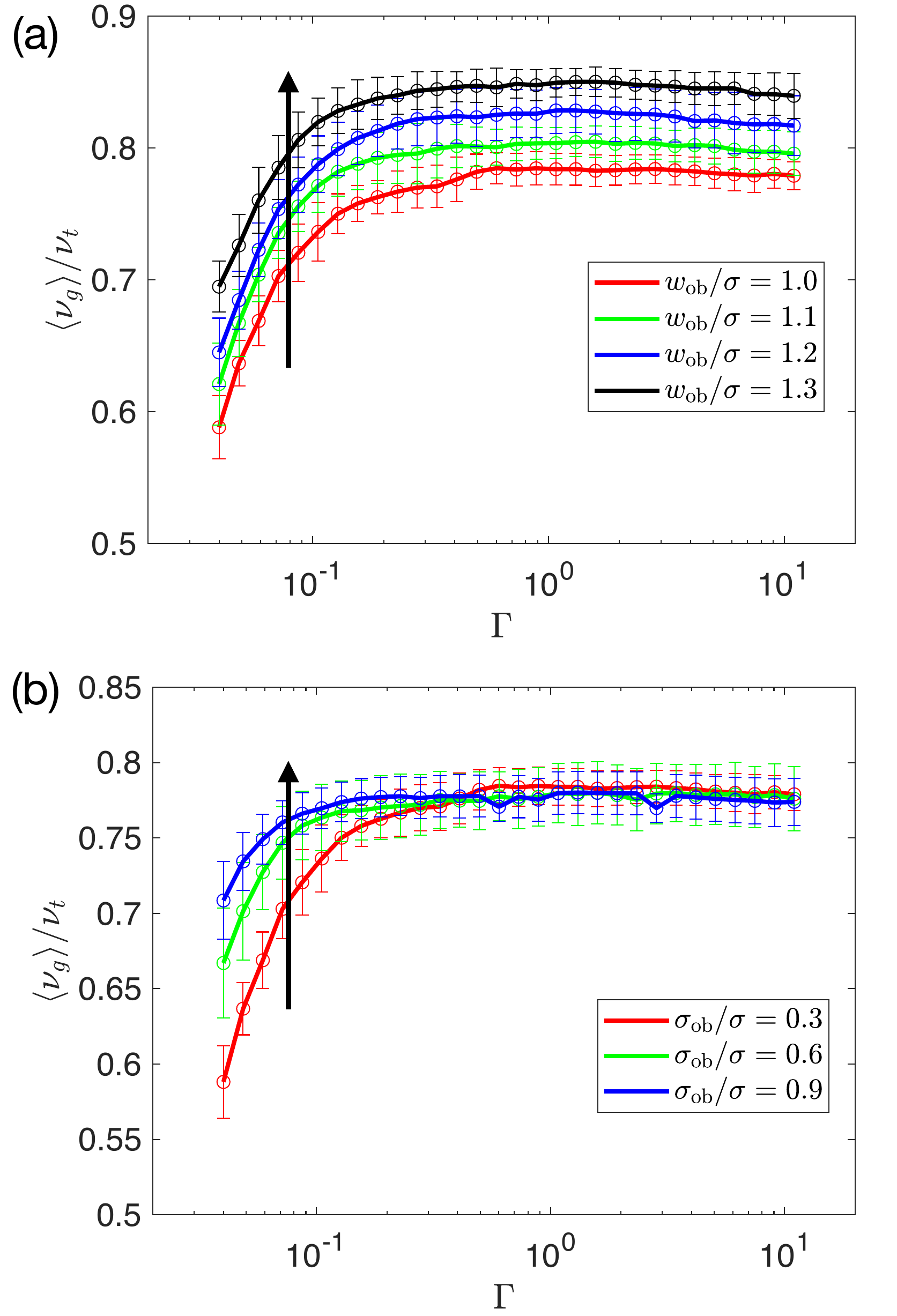}
\centering
\caption{(a) Average speed of the center mass of the droplet in the flow direction $\langle \nu_g\rangle/\nu_t$ plotted as a function of the dimensionless line tension $\Gamma$ for DP simulations of continuous single-droplet flows through obstacle arrays with minimum gap sizes $w_{\rm ob}/\sigma = 1.0$ (red), $1.1$ (green), $1.2$ (blue), and $1.3$ (black). The arrow indicates increasing $w_{\rm ob}/\sigma$. (b) $\langle \nu_g\rangle/\nu_t$ plotted versus $\Gamma$ for DP simulations of continuous single-droplet flows through obstacle arrays with $w_{\rm ob}/\sigma = 1.0$ and $\sigma_{\rm ob}/\sigma = 0.3$ (red), $0.6$ (green), and $0.9$ (blue). The arrow indicates increasing $\sigma_{\rm ob}$.}
\label{fig:arrayvelocity}
\end{figure}

We now investigate continuous flows of single droplets through obstacle arrays (using the SP and DP models). We calculate the average droplet speed $\langle \nu_g \rangle$ (over each droplet trajectory and multiple random obstacle arrays, see Appendix D) as a function of the size and placement of the obstacles. For simplicity, we set $V_t = 0.03$ and $b_0 = 0$, but the results also hold for non-zero $b_0$. (See Appendix E.)  Intuitively, one might expect that increasing the line tension $\Gamma$ would lead to increased resistance to droplet shape changes, and thus decreases in $\langle \nu_g \rangle$. However, in Fig.~\ref{fig:arrayvelocity} (a), we find that $\langle \nu_g \rangle$ increases with $\Gamma$ for $w_{\rm ob}/\sigma \geq 1$. The decrease in $\langle \nu_g \rangle$ with decreasing $\Gamma$ occurs because droplets with small surface tension can easily wrap around an obstacle.  When droplets wrap around an obstacle, the two ends of the droplet extend in the flow direction until the droplet surface energy matches the work done by the driving force. After the ends of droplet stop moving in the flow direction, one end of the droplet moves in the opposite direction of the driving force, which allows the droplet to unwrap. This wrapping and unwrapping process significantly slows droplet movement and decreases $\langle \nu_g\rangle$.

{
Numerous studies have characterized droplet flows as a function of the capillary number $Ca$, which can be expressed in terms of the dimensionless line tension, 
\begin{equation}
\label{ca}
Ca=\frac{b_\infty v_g}{\gamma}=\frac{b_\infty v_g}{b_\infty v_t}\frac{b_\infty v_t}{\gamma}=\frac{v_g}{v_t}\frac{mg}{\gamma}=\frac{v_g}{v_t}\frac{\pi}{4\Gamma}.  
\end{equation}
Since $\Gamma$ and $Ca$ are inversely related, $\nu_g$ versus $Ca$ for $w_{\rm ob}/\sigma \geq 1$ can be inferred from Fig.~\ref{fig:arrayvelocity} (a): $\langle \nu_g \rangle$ is nearly constant with $Ca$ for small $Ca$ and then decreases with $Ca$ at large $Ca$.
}

In Fig.~\ref{fig:arrayvelocity} (b), we quantify continuous flows after fixing $w_{\rm ob}/\sigma=1.0$ and varying $\sigma_{\rm ob}/\sigma$. We plot $\langle \nu_g\rangle/\nu_t$ versus $\Gamma$ for $\sigma_{\rm ob}/\sigma = 0.3$, $0.6$, and $0.9$. We find that increasing $\sigma_{\rm ob}/\sigma$ increases $\langle \nu_g\rangle$ at small $\Gamma$, indicating that it is more difficult for droplets to wrap around larger obstacles. In the large-$\Gamma$ limit, the single-droplet flows for all $\sigma_{\rm obs}/\sigma$ approach the same $w_{\rm obs}$-dependent plateau value in $\langle \nu_g\rangle$. Droplet wrapping does not occur in this regime since droplet deformation would require large surface energy.

\begin{figure}[!h]
\centering
\includegraphics[width=0.95\columnwidth]{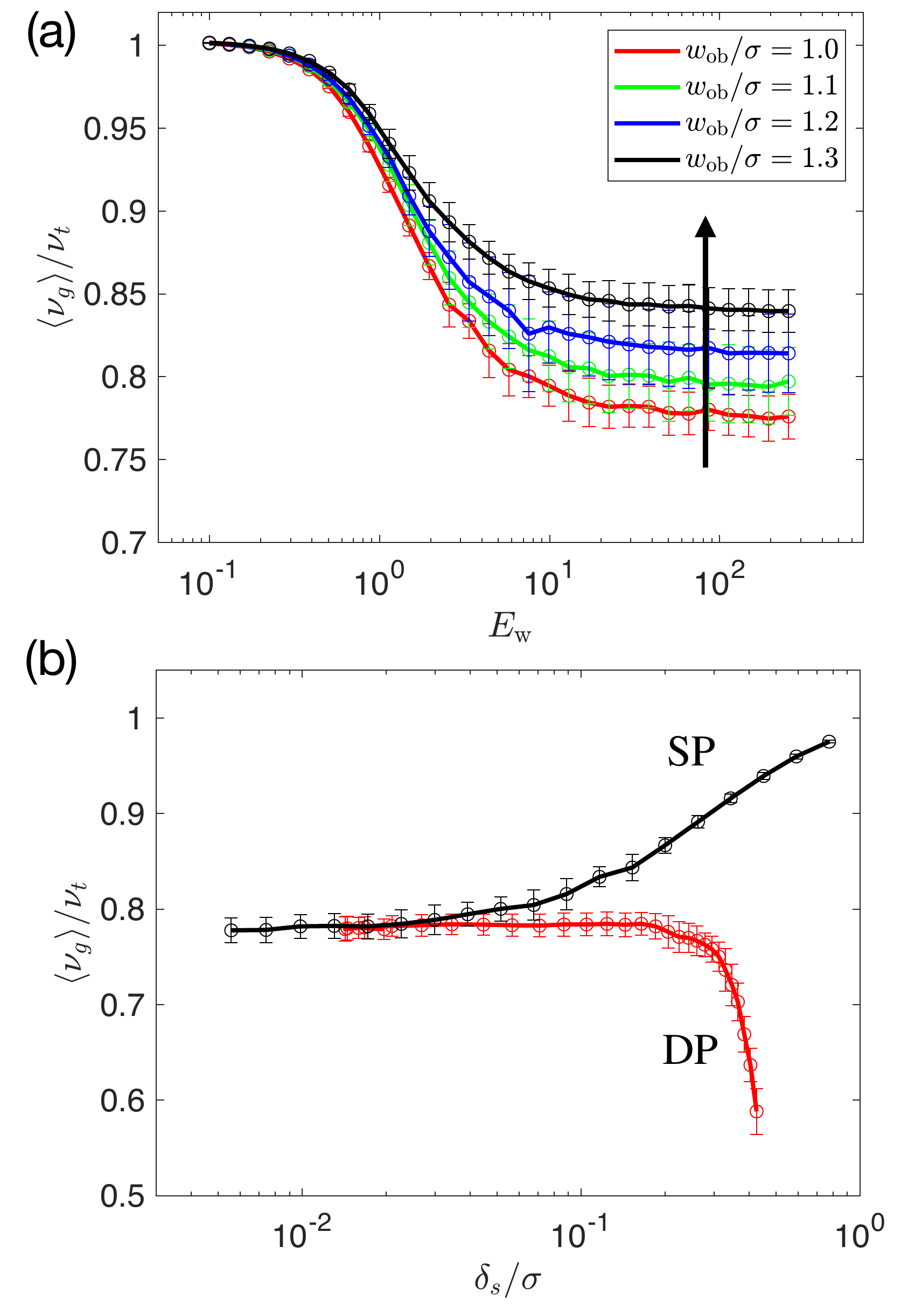}
\centering
\caption{(a) Average speed of the center mass of the droplet in the flow direction $\langle \nu_g\rangle/\nu_t$ plotted as a function of $E_{\rm w}$ for SP simulations of continuous single-droplet flows through obstacle arrays with minimum gap sizes $w_{\rm ob}/\sigma = 1.0$ (red), $1.1$ (green), $1.2$ (blue), and $1.3$ (black). The arrow indicates increasing $w_{\rm ob}/\sigma$. (b) $\langle \nu_g \rangle/\nu_t$ plotted as a function of $\delta_s/\sigma$ for single-droplet flows through obstacle arrays with $w_{\rm ob}/\sigma = 1.1$ for both SP (black) and DP (red) models.}
\label{fig:SP_v_g}
\end{figure}

In Fig.~\ref{fig:SP_v_g} (a), we show $\langle \nu_g \rangle/\nu_t$ versus the droplet-obstacle stiffness $E_{\rm w}$ using the same obstacle arrays in Fig.~\ref{fig:arrayvelocity} for the SP model. In contrast to the DP simulations, we find that $\langle \nu_g \rangle/\nu_t$ decreases with increasing $E_{\rm w}$, which indicates that softer particles flow more rapidly for the SP model. This result stems from the fact that droplet softness in the SP model is represented by larger allowed overlaps between the droplet and obstacles. Therefore, in the $E_{\rm w} \rightarrow 0$ limit, droplets would flow through obstacles without slowing down, resulting in $\langle \nu_g\rangle/\nu_t \rightarrow 1$. In contrast, in the $\Gamma \rightarrow 0$ limit in DP simulations, the droplets wrap around each obstacle that they encounter. This difference in the flow behavior of single SP and DP droplets emphasizes the importance of modeling explicit deformability in droplet flows through obstacle arrays. In Fig.~\ref{fig:SP_v_g} (b), we plot $\langle \nu_g\rangle/\nu_t $ versus droplet deformation $\delta_s/\sigma$ at $w_{\rm ob}/\sigma = 1.1$ for both the DP and SP models. We find that $\langle \nu_g \rangle/\nu_t$ for the SP and DP models converge to same value in the rigid-droplet limit ($\delta_s/\sigma \rightarrow 0$), but they possess diverging behavior for large droplet deformation, $\delta_s/\sigma \rightarrow 1$.


\section{Discussion and Conclusions}
\label{discussion}

In this article, we carried out coordinated experiments and numerical simulations of gravity-driven flows of single droplets in narrow channels and obstacle arrays. We first developed deformable particle (DP) simulations with line tension in 2D and calibrated them to the experiments. The DP simulations recapitulate the experimental results for the droplet shape and speed for flows within narrow channels with an error of less than $10\%$. Using the calibrated DP simulations, we find several important results concerning single-droplet flows in narrow channels and obstacle arrays. First, the droplet speed possesses nonmonotonic behavior as the droplet traverses the narrow channels. The droplet speed can even overshoot the terminal speed and the overshoot is enhanced as the gravitational driving is decreased.  Second, we showed that the clogging probability is nonmonotonic with the line tension for droplets flowing through obstacle arrays. The nonmonotonic behavior is caused by two distinct clogging mechanisms (droplet wrapping and squeezing) in obstacle arrays. Stiffer droplets become trapped when they cannot squeeze in between obstacles, while highly deformable droplets clog when they wrap around obstacles, resulting in high clogging probabilities for droplets with both large and small deformabilities. We also compared the DP simulation results for flows in obstacle arrays to those for the frequently used soft particle (SP) model, which does not include explicit shape deformability. The DP and SP simulations of single-droplet flows in obstacle arrays are similar in the rigid-droplet limit, but diverge for deformable droplets since the SP model can only model clogging via the squeezing mechanism (and not the wrapping mechanism).

These results suggest several promising future research directions. First, the current numerical simulations employ the DP model with line tension in 2D. When calibrating the 2D DP model to the experimental results, we obtain values for the dimensionless line tension $\Gamma^*$ that are a factor of $\sim 3$ smaller than the dimensionless surface tension $\Gamma_{\rm exp}$ in experiments. Thus, in future work we will develop the 3D DP model~\cite{Wang2021} with surface tension to describe quasi-2D flows of droplets in narrow channels and obstacle arrays. Second, In the current work, we do not quantitatively model the background fluid. However, the hydrodynamics of the background fluid can influence the droplet dynamics, such as causing the droplets to follow stream lines, which can prevent droplet-obstacle contact. Therefore, in the future, we can include the effects of Stokesian dynamics of the background fluid on the droplet motion. 

Third, we observe in experiments that droplets undergoing sufficiently large deformations (larger than those considered in the results section) break up into smaller droplets. However, our current numerical simulations do not account for this phenomenon. In future studies, we will develop a quantitative framework to predict the conditions under which droplet breakup occurs and use it to determine the size distribution of daughter droplets emerging from obstacle arrays. This improved understanding of droplet breakup will be valuable in microfluidic applications \cite{PhysRevLett.92.054503, CHUNG200938}, where generating droplets of a specific size is required \cite{TazikehLemeski2024}. We also plan to investigate multi-droplet flows through obstacle arrays. The efficiency of the DP model will enable us to simulate systems consisting of thousands of droplets, and explore under what conditions droplet interactions give rise to coalescence, where two droplets in contact merge to form a larger droplet \cite{PhysRevE.104.014702}. Droplet breakup and coalescence play competing roles in determining the equilibrium size distribution of droplets moving through obstacle arrays \cite{BARAI20191,DENG2024104953}. In future studies, we will investigate their combined effects to quantitatively predict the droplet size distribution exiting from the obstacle arrays.

{Fourth, we observe cases where both wrapping and squeezing mechanisms occur simultaneously during droplet flows. These hybrid cases occur at intermediate values of the line tension, i.e. where the wrapping and squeezing branches connect in Fig.~\ref{fig:clogp} (c). The occurrence of hybrid behavior also depends on the obstacle density. For example, at sufficiently high density, a droplet could unwrap around one obstacle and the side of the droplet that is unwrapping can become squeezed between two obstacles that are further downstream. In the future, we will investigate the conditions that cause these hybrid cases and their effects on the flow and clogging of capillary droplets.}

{Finally, the droplet squeezing and wrapping mechanisms can be exploited in deterministic-lateral-displacement (DLD) devices to separate droplets spatially and/or temporally based on their surface tension. Currently, DLD devices separate droplets based on their size and response to fluid streamlines.  Our results pave the way for separating droplets or cells in DLD devices at fixed size but varying surface tension.}


\begin{figure}[!h]
\centering
\includegraphics[width=0.99\columnwidth]{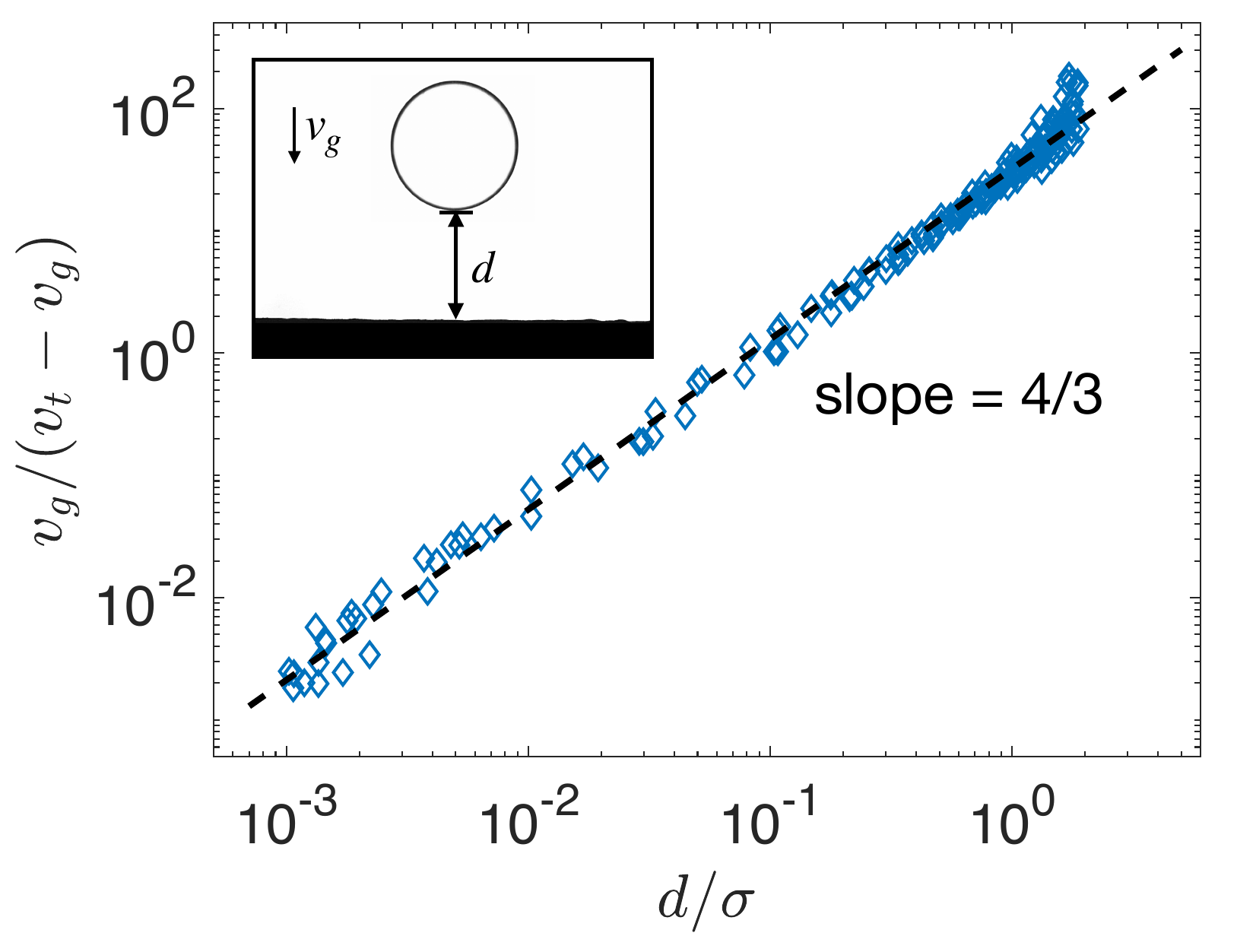}
\centering
\caption{The ratio $\nu_g/(\nu_t - \nu_g)$ plotted versus distance $d/\sigma$ from the chamber wall for the experiments. The droplet has diameter $\sigma = 0.98 ~\rm mm$, tilt angle $\theta = 5^\circ$, and terminal velocity far away from the wall, $v_t = 0.65 ~\rm mm/s$. An image from the experiment is shown in the inset. The dashed line has slope $4/3$.}
\label{fig:drag}
\end{figure}

\section*{Appendix A}
\label{app:A}

In this Appendix, we show experimental measurements of the power-law exponent $\alpha$ that appears in Eq.~\ref{eq:drag} for the distance-dependent drag coefficient. In the low Reynolds number (overdamped) regime, Eq.~\ref{eq:drag} can be rewritten as:
\begin{equation}
\label{intermediate}
\nu_g \zeta = \nu_g b_{\infty}\left(1 + \frac{b_0}{b_{\infty}} \left(\frac{\sigma}{d}\right)^{\alpha}\right) \approx -M g = \nu_t b_{\infty}.
\end{equation}
After rearranging Eq.~\ref{intermediate}, we find
\begin{equation}
b_0 \left(\frac{\sigma}{d}\right)^{\alpha} \approx \left(\frac{\nu_t}{\nu_g} - 1\right) b_{\infty},
\end{equation}
\begin{equation}
\alpha \log_{10} \left(\frac{d}{\sigma}\right) \approx \log_{10}\left(\frac{\nu_g}{\nu_t - \nu_g}\right) + \log_{10}\left( \frac{b_{\infty}}{b_0} \right).
\end{equation}
In Fig.~\ref{fig:drag}, we plot $v_g/(v_t - v_g)$ versus $d/\sigma$ on a $\log_{10}$-$\log_{10}$ scale, where $d$ is the minimum distance between chamber wall and the droplet. The droplet shown in the inset has diameter $\sigma = 0.98 ~\rm mm$, tilt angle $\theta = 5^\circ$, and terminal velocity far from wall $v_t = 0.65 ~\rm mm/s$. We find that the experimental data in Fig.~\ref{fig:drag} is well-fit by a power-law exponent $\alpha =4/3$. 

\section*{Appendix B}
\label{app:B}

\begin{figure}[!h]
\centering
\includegraphics[width=0.99\columnwidth]{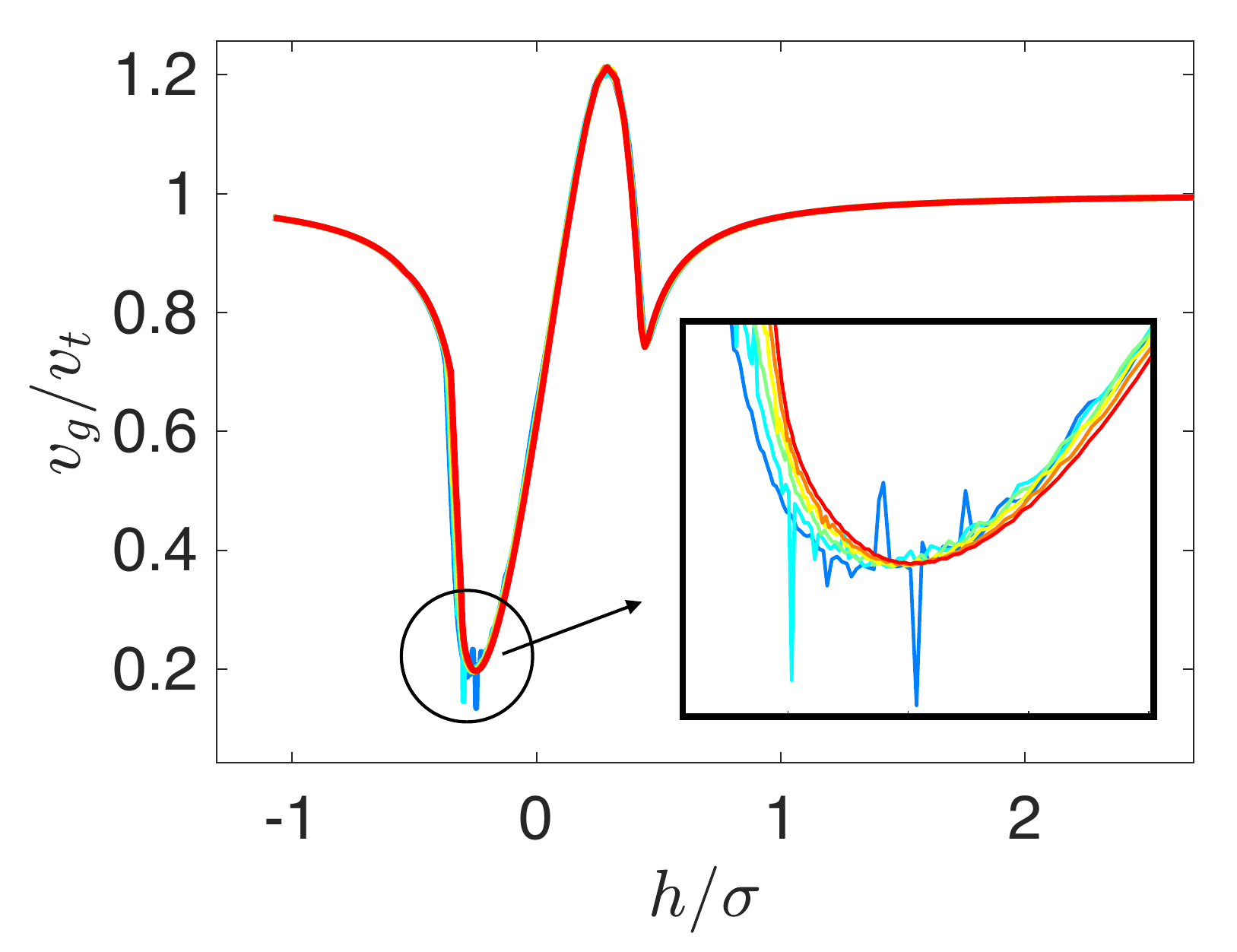}
\centering
\caption{Droplet center of mass speed $\nu_g/\nu_t$ plotted as a function of distance $h/\sigma$ from the orifice for DP simulations of a droplet flowing through a narrow channel with $w=0.7\sigma$, $\Gamma = 1.08$, and $N_v = 32$, $64$, $128$, $256$, $512$, and $1024$. The inset is a close-up of $\nu_g/\nu_t$ near the first minimum in the speed at $h/\sigma \approx -0.25$ to highlight the results for varying $N_v$, from $32$ (blue) to $1024$ (red).}
\label{fig:NV}
\end{figure}

In this Appendix, we investigate the effects of the number of vertices $N_v$ and perimeter elasticity $K_l$ on the results for droplet flows through narrow channels using the DP simulations. The DP model with $N_v \rightarrow \infty$ and $K_l \rightarrow 0$ is the most accurate for describing droplet flows. However, to limit computational time, we choose $N_v$ such that it is sufficiently large that further increases do not significantly change the results. In Fig.~\ref{fig:NV}, we show the droplet speed profile, $\nu_g/\nu_t$ versus $h/\sigma$, for $N_v = 32$, $64$, $128$, $256$, $512$, and $1024$. We find that for $N_v \le 64$, the velocity profile is quite noisy. For $N_v \ge 128$, the velocity profile is smooth on a linear scale. For the DP simulations presented here, we use $N_v = 256$.

\begin{figure}[!h]
\centering
\includegraphics[width=0.9\columnwidth]{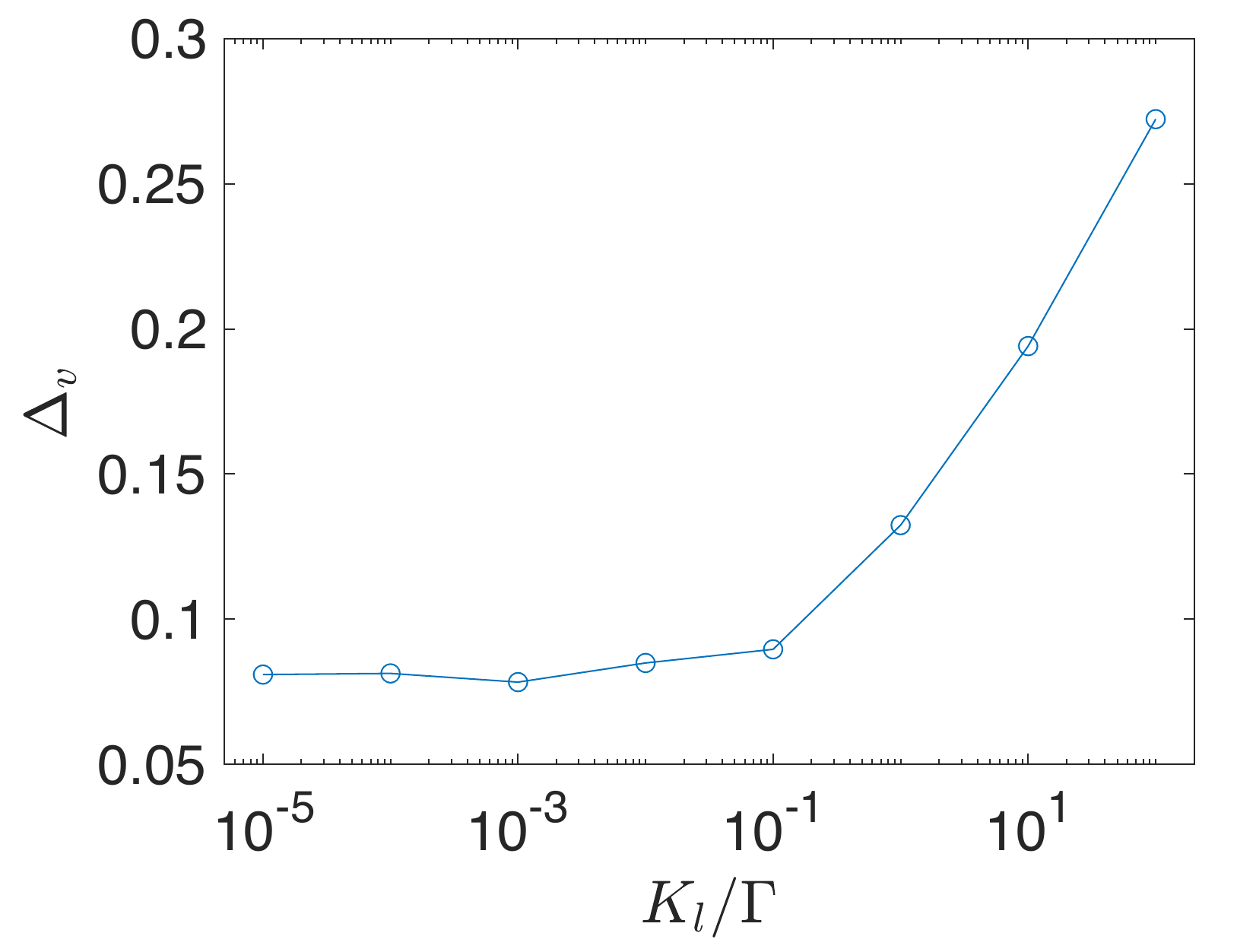}
\centering
\caption{Root-mean-square deviation in the droplet speed $\Delta_{\nu}$ between the DP simulations and experiments for a droplet flowing through a narrow orifice with width $w/\sigma=0.9$ plotted as a function of $K_l / \Gamma$. $\Delta_{v} \sim 0.08$ reaches a plateau for $K_l/ \Gamma < 10^{-2}$.}
\label{fig:Kl}
\end{figure}

For the DP simulations to accurately model line tension, it is important that $K_l/\Gamma \rightarrow 0$. However, the perimeter elasticity in Eq.~\ref{shape_energy} is important for ensuring the stability of the DP simulations.  If the DP model only includes line tension without perimeter elasticity, the droplet vertices can aggregate in the direction of gravity, which gives rise to an uneven distribution of vertices around the droplet perimeter and can cause the droplet to overlap the side walls. Thus, we need to select a non-zero value of $K_l$ to uniformly distribute vertices on the droplet surface, but small enough such that the results for the DP model are dominated by line tension and do not depend on $K_l$.  In Fig.~\ref{fig:Kl}, we show the root-mean-square deviations in the droplet speed $\Delta_{v}$ between the DP simulations and experiments of flow in narrow channels plotted as a function of $K_l/\Gamma$. $\Delta_v \sim 0.08$ reaches a plateau for $K_l/\Gamma \lesssim 10^{-2}$. In the present work, we set $K_l/\Gamma =10^{-3}$.

\section*{Appendix C}
\label{app:E}
\begin{figure}[!h]
\centering
\includegraphics[width=0.9\columnwidth]{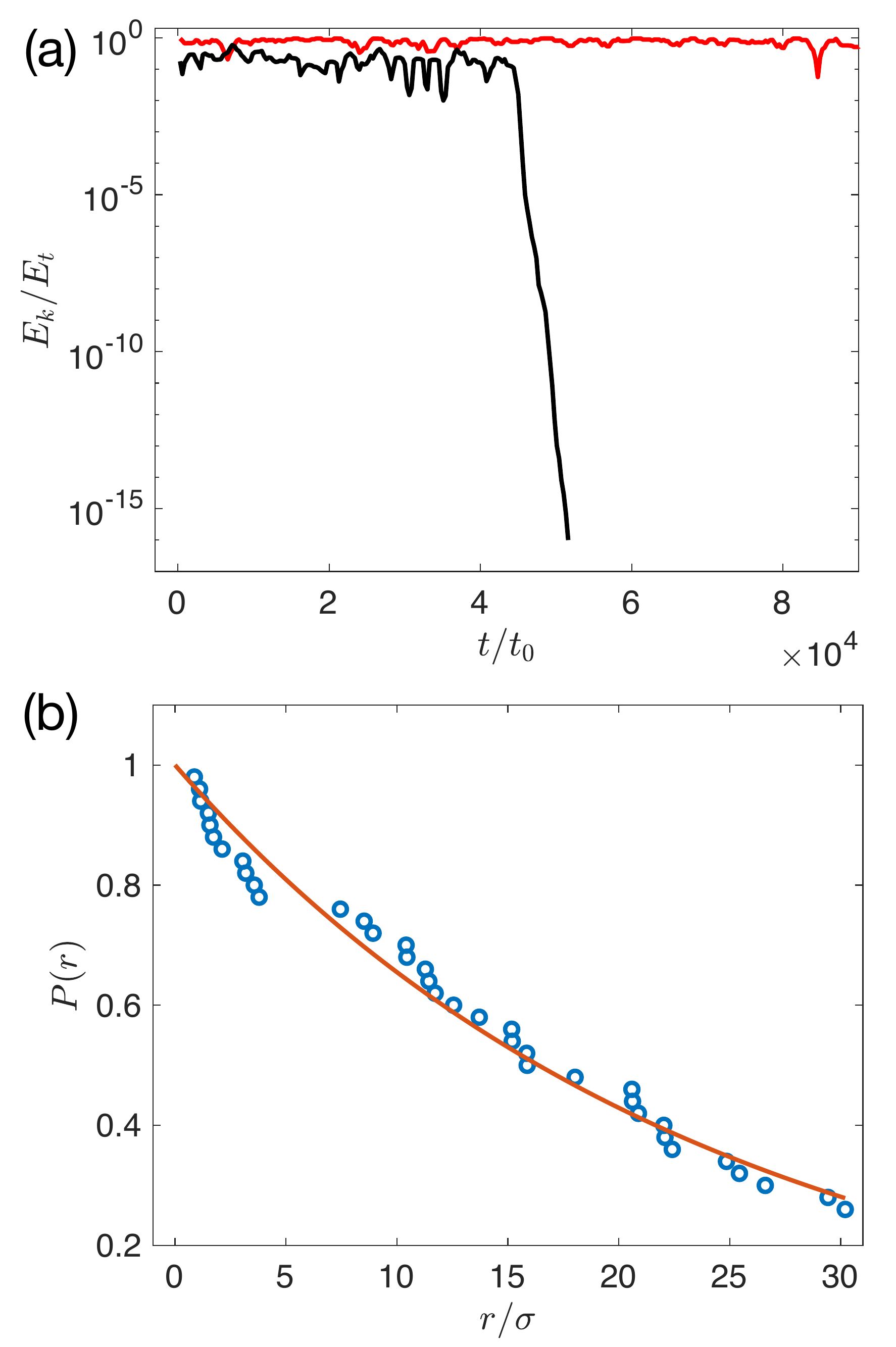}
\centering
\caption{(a) Droplet kinetic energy $E_k$, normalized by the kinetic energy at the terminal speed $E_t$, plotted as a function of time $t/t_0$ for droplets moving in obstacle arrays with $\Gamma = 0.5$, $\sigma_{\rm ob}/\sigma = 0.3$, and both $w_{\rm ob}/\sigma = 1.1$ (red solid line) and $0.3$ (black solid line). For $w_{\rm ob}/\sigma = 0.3$, $E_k/E_t$ decays exponentially toward the clogging threshold $10^{-17}$. For $w_{\rm ob}/\sigma = 1.1$, the droplet flows continuously without clogging. (b) Cumulative probability $P(r)$ of the droplet not forming a clog plotted as a function of the displacement in the obstacle array $r/\sigma$ using DP simulations with $\Gamma \approx 1.09$, $\sigma_{\rm ob}/\sigma = 0.3$, and $w_{\rm ob}/\sigma = 0.2$. The red solid curve represents $P = \exp(-r/\lambda)$, where $\lambda/\sigma = 24 \pm 1$ is the clogging decay length. }
\label{fig:Ek}
\end{figure}

In this Appendix, we include additional details concerning the definition of a permanent clog and the clogging decay length $\lambda$ for droplets moving in obstacle arrays. In the simulations, a clog occurs when the droplet kinetic energy $E_k$ (normalized by the kinetic energy at the terminal speed) decays below a threshold $E_k/E_t < 10^{-17}$. In Fig.~\ref{fig:Ek} (a), we show examples of $E_k/E_t$ versus time $t/t_0$ for cases when the droplet clogs ($w_{\rm ob}/\sigma = 0.3$) and does not clog ($w_{\rm ob}/\sigma = 1.1$). 

To determine the clogging decay length $\lambda$, we first measure the cumulative probability $P$ that the droplet does not clog as a function of the droplet displacement $r$ in the obstacle array. We run $N_s = 50$ simulations for a total time $T = 2 \times 10^5 t_0$ with different randomized initial droplet positions and randomized obstacle arrays. $P(r) =1 -  N_{\rm clog}(r)/N_s$, where $N_{\rm clog}$ is the number of times that the droplet clogs before it moves a distance $r$. In Fig.~\ref{fig:Ek} (b), we plot $P$ versus $r/\sigma$ for DP simulations with $\Gamma \approx 1.09$, $\sigma_{\rm ob}/\sigma = 0.3$, and $w_{\rm ob}/\sigma = 0.2$. At the beginning of each simulation, $r = 0$ and $P = 1$. As each droplet traverses the obstacle array, the cumulative probability $P$ of not clogging decreases exponentially with increasing $r$.

\section*{Appendix D}
\label{app:C}

\begin{figure}[!h]
\centering
\includegraphics[width=0.9\columnwidth]{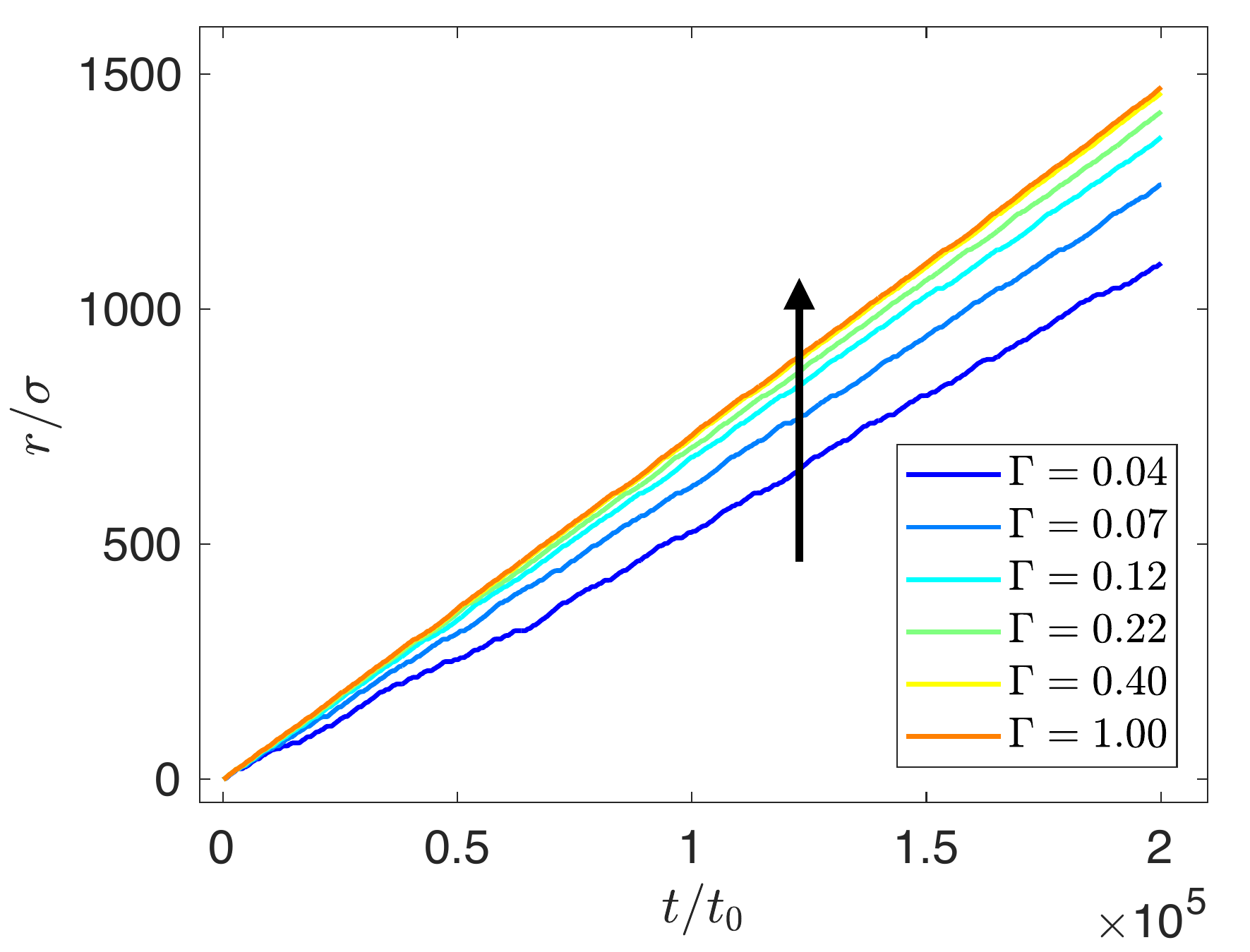}
\centering
\caption{Droplet displacement $r$ in the flow direction $r/\sigma$ plotted as a function of time $t/t_0$ as the droplet traverses an obstacle array with $w_{\rm ob}/\sigma=1.1$ and $\sigma_{\rm ob}/\sigma = 0.3$. The arrow indicates increasing $\Gamma$ from $0.04$ (blue) to $1$ (red).}
\label{fig:XvsT}
\end{figure}

In this Appendix, we show that the DP simulations of single droplet flows in obstacle arrays can reach a continuous flow regime (with no permanent clogs) for minimum obstacle spaceing $w_{\rm ob}/\sigma > 1$. In Fig.~\ref{fig:XvsT}, we plot the droplet displacement $r$ versus time $t$ for DP simulations with $\Gamma = 0.04$-$1$. We determine the steady state speed $\nu_g$ for a single-droplet trajectory in a given obstacle array by measuring the slope of $r/\sigma$ versus $t/t_0$. $\langle \nu_g\rangle$ is averaged over $10$ random obstacle arrays.

\section*{Appendix E}
\label{app:D}
\begin{figure}[!h]
\centering
\includegraphics[width=0.9\columnwidth]{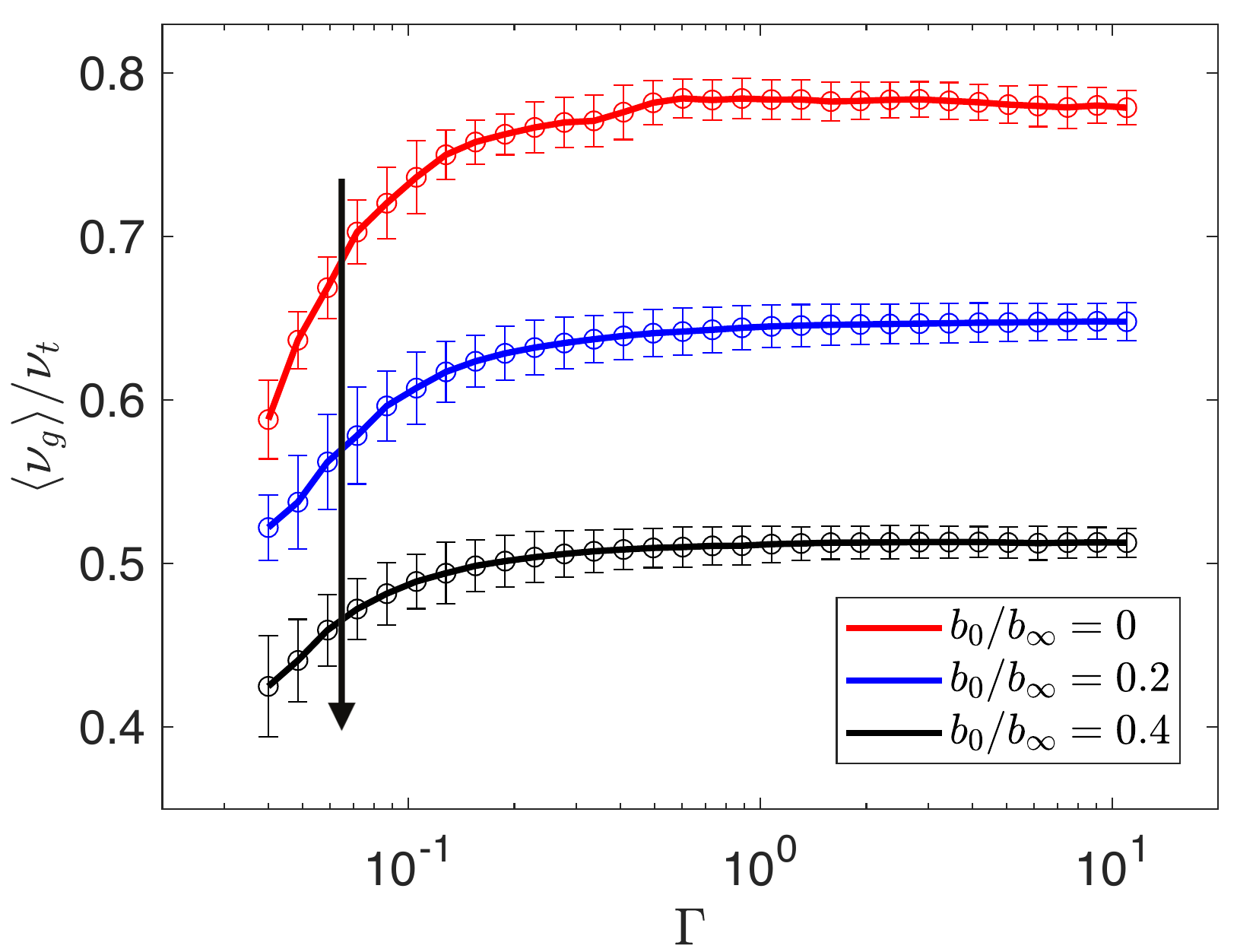}
\centering
\caption{Average speed of the center mass of the droplet in the flow direction $\langle \nu_g\rangle$ normalized by the terminal velocity $\nu_t$ plotted as a function of the dimensionless line tension $\Gamma$ for obstacle arrays with $w_{\rm ob}/\sigma = 1.0$. All other parameters are the same as those used in Fig.\ref{fig:arrayvelocity} (a), except we vary the near-wall drag coefficient: $b_0/b_{\infty}= 0$ (red), $0.2$ (blue), and $0.4$ (black). The arrow indicates increasing $b_0/b_{\infty}$.}
\label{fig:b0}
\end{figure}

In this Appendix, we show that the choice of the near-wall drag coefficient $b_0$ does not qualitatively affect the results in Sec.\ref{sec:v_g}. As discussed previously, the SP model does not include distance-dependent drag forces (Eq.\ref{eq:drag}). In Fig.~\ref{fig:SP_v_g} (c) and Fig.\ref{fig:SP_p} (b), we set $b_0=0$ in the DP simulations so that when we compare the DP and SP simulations, both have the same form for the drag force. In Fig.~\ref{fig:b0}, we plot $\langle \nu_g\rangle/\nu_t$ versus $\Gamma$ for single droplets flowing through obstacle arrays with $w_{\rm ob}/\sigma = 1.0$. All other parameters are the same as those in Fig.\ref{fig:arrayvelocity} (a), except the near-wall drag coefficient varies: $b_0/b_{\infty}= 0$, $0.2$, and $0.4$. We find that increasing $b_0/b_{\infty}$ decreases $\langle \nu_g\rangle/\nu_t$, but it monotonically increases with $\Gamma$ for all $b_0/b_{\infty}$. 

\section*{Acknowledgments}
We acknowledge support from NSF Grants No. CBET-2002782 (Y. C., S.S., and C. S. O.), No. CBET-2002815 (B. L., P. H., and E. R. W.), No. CBET-2306371 (D. M.), and No. CBET-2002797 (M. D. S.). This work was also supported by the High Performance Computing facilities operated by Yale’s Center for Research Computing.

\bibliography{bib}
\bibliographystyle{rsc}
\end{document}